\documentclass{sig-alternate} 
\usepackage{mathptmx} 
\usepackage[T1]{fontenc}
\usepackage[utf8]{inputenc}
\usepackage{fancyhdr}
\usepackage[normalem]{ulem}
\usepackage[hyphens]{url}
\usepackage{microtype}
\usepackage{algorithm}
\usepackage{algpseudocode}
\usepackage{xcolor}
\usepackage{pifont}

\newcommand{\ignore}[1]{}

\usepackage[bookmarks=true,breaklinks=true,letterpaper=true,colorlinks,linkcolor=black,citecolor=blue,urlcolor=black]{hyperref}

\newcommand{\needcite}[1]{{\color{red}[] }}

\fancypagestyle{firstpage}{
  \fancyhf{}

  \fancyhead[C]{} 
  \fancyfoot[C]{\thepage}
}  

\title{Memory Slices: A Modular Building Block for Scalable, Intelligent Memory Systems\vspace{-2.5em}} 

\numberofauthors{3} 
\author{
\alignauthor
Bahar Asgari\\
       \affaddr{Georgia Institute of Technology}\\
\alignauthor
Saibal Mukhopadhyay\\
       \affaddr{Georgia Institute of Technology}\\
\alignauthor 
Sudhakar Yalamanchili\\
       \affaddr{Georgia Institute of Technology}\\
}

\begin{document}
\maketitle
\thispagestyle{firstpage}
\pagestyle{plain}

\begin{abstract}

 While reduction in feature size makes computation cheaper in terms 
of latency, area, and power consumption, performance of emerging 
data-intensive applications is determined by  
data movement, which is becoming more costly.
These trends have introduced the concept of scalability as reaching a 
desirable performance per unit cost by using as few number of units as 
possible. 
Therefore, many proposals have moved compute closer to the
memory, used modular and parallel architectures, and utilized 
interconnection networks. 
However, these efforts ignored maintaining a balance between 
bandwidth and compute rate of an architecture, with those of applications, 
which is a key principle in designing scalable large systems. 
This paper proposes the use of {\em memory slices}, a modular building block 
for scalable memory systems integrated with compute, in which performance 
scales with memory size (and volume of data). 
The slice architecture utilizes a programmable memory interface feeding a
systolic compute engine with high reuse rate. 
The modularity feature of slice-based systems is exploited with a 
partitioning and data mapping strategy across allocated memory slices where
training performance scales with the data size.
These features enable shifting the most pressure to cheap compute units 
rather than expensive memory accesses or transfers via interconnection network.
An application of the memory
slices to a scale-out memory
system is accelerating the training of recurrent, convolutional, and hybrid 
neural networks (RNNs and RNNs+CNN) that are forming cloud workloads. 
The results of our cycle-level simulations show that memory slices exhibits 
a superlinear speedup when the number of slices increases. 
Furthermore, memory slices improve power efficiency to 747 GFLOPs/J for
training LSTMs.
While our current evaluation uses memory slices with 3D packaging, a major 
value is that slices can also be constructed with a variety of packaging
options, for example with DDR-based memory units.

\end{abstract}

\section{Introduction}
\noindent

As feature size of CMOS technology scales down, the cost of a billion floating 
point (GFLOPs) decreases at a rate of 35\% operations per year~\cite{dally2008digital}. 
Moreover, every five years, floating point units (FPUs) can provide 8X arithmetic operations per 
Watt, and similarly, FPUs in a given area perform 8X faster with the 
same cost \cite{dally2003merrimac}. 
However, the bandwidth cost is increasing with distance~\cite{dally2008digital}. 
As a result, computing is at an inflection point where the escalating energy and
latency costs of data movement are dominating that of
computation \cite{keckler2011gpus, borkar2013role, mai2000smart, 
dally2003merrimac}. 
This has been amplified by the exponential growth of data
and the explosive growth in new algorithmic approaches such as machine
learning, graph analytics, relational processing, and stochastic
programming to convert these exponentially growing data sets into useful information. 
This rising impact of data movement has spawned
many proposals for moving compute closer to the memory system in the
form of accelerators to reduce the data movement costs\cite{gao2015practical, 
farmahini2015nda, asghari2016chameleon, nai2017graphpim, kim2016neurocube, gao2017tetris, 
o2017fine}, which have included integrating compute logic onto the memory die as well as 
into the package, (e.g., DIMM, interposer, or 3D die stacks). 
 In addition, studies such as high-bandwidth DRAM architectures (e.g., FGDRAM) 
~\cite{o2017fine} for increasing memory bandwidth at low energy, modular architectures 
that utilize efficient interconnection networks for 
shrinking wire length ~\cite{mai2000smart}, and stream processors for
reducing the average distance that an operand must travel to reach an 
FPU~\cite{dally2003merrimac} have been proposed contributions for dealing with scaling 
issues.

 An un-covered factor in providing scalability by studies that leveraged 
on-chip SRAM with integrated DRAM stacks connected to GPUs, FPGAs, TPUs, or any other 
compute unit, is maintaining the matched balance between compute rate and 
bandwidth, which in turn enables achieving a given desirable performance with fewer number 
of nodes. 
This paper addresses this gap by proposing {\em Memory Slices} - a modular building
block for constructing scalable memory systems integrated with compute
logic. Inspired by the bit-slice microprocessors~\cite{kanttaiah1978bit} of earlier decades, a
memory slice is a unit of memory integrated with compute logic and an
on-chip switched interconnection network (ICN) interface. Slices are
aggregated into large memory systems wherein local memory bandwidth to
the compute logic scales linearly with compute bandwidth, and external
bandwidth grows as a function of the number of slices and can be
delivered through memory networks such as proposed
in~\cite{kim2013memory}. 

Slice memory controllers are augmented with
programmable address generators for traversing memory data structures
creating a data driven execution model for the compute logic. The
specific slice microarchitecture proposed in this paper uses a
pipelined, systolic-style compute engine~\cite{kung1988vlsi, gentleman1982matrix} 
to balance local slice memory
bandwidth with slice compute bandwidth. The collective fine-grained
distributed compute logic across slices can host complex computations
{\em in the memory system} on large data sets leading to the notion of
intelligent memory – for example when acting as a host for machine
learning algorithms.
Moreover, the fine-grained implantation is not only highly efficient for 
applications with varied data sizes in their steps (e.g. neural networks), but also is 
compatible with fine-grained packet-switched low-power memories.

This paper reports on the application of the memory slice
microarchitecture to the scale-out organization of memory systems to
accelerate an important class of deep neural networks (DNNs) namely
recurrent neural networks (RNNs) and hybrid neural networks composed
of convolutional neural networks and RNNs (CNN+RNN). 
This class of on-demand applications are not only dense but also require high rate 
of reuse (i.e., high ratio of FLOPs:Byte). They also include many independent operations of 
same type. Because of these features, they favors fine-grained parallel architectures, with 
minimum distance (no hierarchy of registers) between memory and FPUs, with high reuse rate, 
which is arranged in memory slice architecture.
Note that while the majority of past efforts have focused on the acceleration of CNNs and mainly
focused on supporting inference with training performed off-line over
GPU clusters~\cite{parashar2017scnn, gao2017tetris, kim2016neurocube,
  chen2017eyeriss, chen2014diannao, chen2014dadiannao,
  liu2015pudiannao, du2015shidiannao}. However, RNNs represent an
emerging class of DNNs for learning temporally dependent patterns and
are becoming the dominant workloads in the
cloud~\cite{jouppi2017datacenter}. Training of RNNs has unique
computational characteristics that challenges acceleration, presents
substantial data movement management challenges, and therefore impedes
progress towards expanding its scope of application to future larger
applications. 

This paper shows how the modular microarchitecture of
slices are exploited in the partitioning and mapping of data where
training performance can scale with the size of data and at much lower
energy than GPU clusters. For example, aggregation functions operate
transparently across slices via the inter-slice network and can
naturally encapsulate activation functions.  Importantly, the scalable
implementation of fine grained parallel dense matrix multiplication is
a core kernel in the implementation described here. When viewed in
this manner, we envision wider application to of applications in
machine learning, graph analytics, and scientific
computations. Mapping applications is now akin to memory allocation
and data layout in memory memory systems and is key to scaling performance
in proportion to size of memory.

The results of our cycle-level simulations show that in average training CNNs on memory slices
consisting of 3D DRAM stacks and integrated computation logic provides 6.3X as high throughput (images/sec) as
Tesla\textsuperscript{\textregistered}P100 GPUs do. 
Moreover, our evaluation exhibits a superlinear speedup when
the number of slices increases from 2 to 256. 
While our current evaluation uses memory slices with 3D packaging, the major value is
that slices can be constructed with a variety of alternative packaging
options including conventional DDR (e.g., to form intelligent DIMMs),
High Bandwidth Memory (HBM)~\cite{standard2013high}, Hybrid Memory
Cube (HMC)~\cite{hybrid2013hybrid},or other emerging memory standards.
Memory slice seeks to make the following key contributions:
\begin{itemize}
\item It is a modular in-memory building block of scalable acceleration of data-intensive applications,
      which shifts the most stress to cheap computations, located near data rather than accessing memory. 
\item It tries to provide a best balance between the bandwidth and compute rate of architecture, with those of applications, 
      to gain a desirable performance at lower cost (fewer nodes). This
      is achieved by exploiting the modular nature of a slice with a flexible data
      partitioning and mapping strategy supported by a data driven
      execution model.
\item It is comprised of i) a systolic compute engine, ii) programmable memory address
      generator, iii) wormhole switched network interface, and iv)
      aggregation engines, all of which are arranged to reach the key goal of scalability.
\item It demonstrates scaling performance with memory
      system size for an on-demand class of dense and compute intensive problems, 
      training of  RNNs and hybrid networks (CCNs + RNNs).
\end{itemize}

\section{Motivation}
\begin{figure}[t]
\centering
\includegraphics[width=1\linewidth]{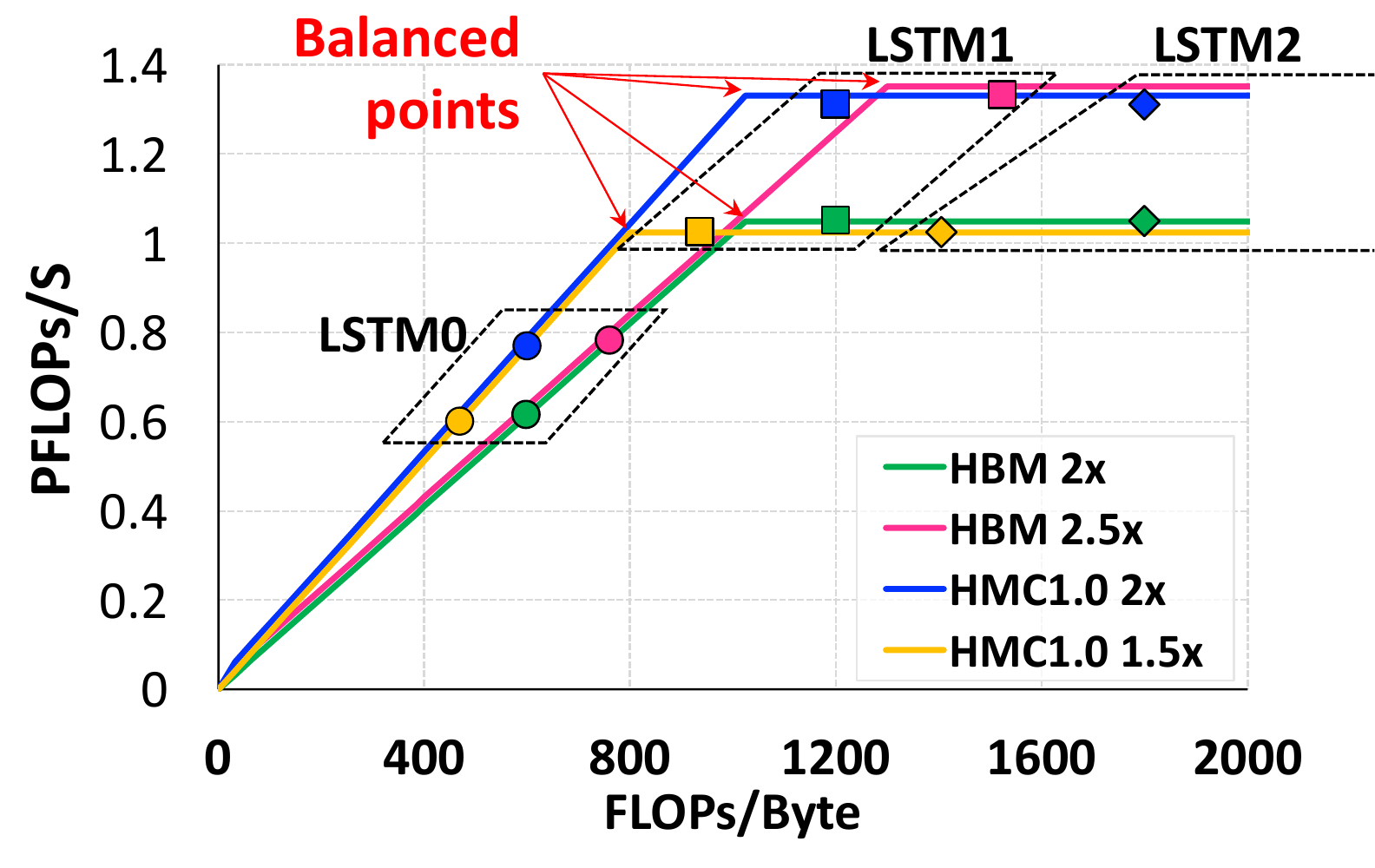}
\vspace{-30pt}
\caption{The Roofline model: the maximum attainable throughput for training three LSTM networks (see Table \ref{Cfg} for configuration details)}
\label{figRoofLine}
\vspace{-0.1in}
\end{figure}

\noindent

Scaling performance of traditional parallel-computing have been challenging 
mainly because of their power budget limitations and using slow-growing memory bandwidth in 
combination of fast-growing computational units which creates a mismatch between 
computation rate and memory bandwidth and leads to in-efficiency. These challenges, which have not 
been design mistakes, worked well for traditional applications. However, they cannot 
meet the requirements of a category of new dense applications with high ratio of FLOPs:byte (i.e. 
reuse).
As a result, the primary motivation of this work is to conceive a near data processing system
architecture wherein the performance scales with the size of memory
(and hence volume of data). From our perspective, this view has
several consequences. First, it implies the maintenance of a fixed
ratio of compute throughput per unit of memory as the size of memory
increases. If we double the size of memory, the compute capacity also
doubles. Second, to scale performance implies that memory bandwidth to
the local compute logic should not reduce with system size (i.e., at
least remain at a fixed ratio relative to the throughput of the
local compute logic).  Therefore, as memory size scales, so does the local
memory bandwidth to compute logic. Less obvious, is the implication
that this leads to the design of bandwidth-optimized memory systems
rather than capacity-optimized systems we have today, where the memory
bandwidth delivered to cores is 2-3 orders of magnitude less than
available within the memory die. 
With the compute logic integrated in the memory system, maximizing
performance is pursued by maximizing fine-grained (word-level)
concurrency.  Data parallel algorithmic implementations that can
maximally expose fine-grained concurrency across large data structures
can be exploited to scale performance with memory size. In this paper,
we propose and evaluate a data-driven, streaming model of computing
within and across memory slices in an attempt to maximize fine-grained
concurrency.

Finally, we wish to demonstrate the proposed approach with
applications to the in-memory acceleration of an important class of
computationally intensive, data parallel applications. We apply this
architectural approach to specific class of state-of-the-art DNN,
namely RNNs and hybrid neural networks (RNNs + CNNs.)  RNNs are
on-demand networks used in applications such as speech recognition or
machine translation in which the data is a sequence. For example, in a
machine translation application, a neuron, which can be a gated
recurrent unit (GRU)~\cite{cho2014properties}, or a long short-term
memory (LSTM)~\cite{hochreiter1997long} processes a sequence of words
and thus operates over temporal sequences. It has been observed that
such workloads are forming the bulk of server class workloads (rather
than the more popularly analyzed CNNs). Training of such networks
involves computationally intense kernels dominated by dense matrix
operations. We wish to demonstrate that memory slice designs can
provide scalable acceleration for such problems, which today are
largely pursued using GPU clusters. We will show that our solution
realizes some unexpected advantages over the GPU clusters from the
perspective of scaling.

The definition of the architecture of a memory slice involves
balancing local memory bandwidth and the throughput of the associated
compute logic. The desired balance is determined by the properties of
the applications/algorithms to be hosted in the memory system. We can
gain some insights as to appropriate balance by examining the Roofline
model\cite{williams2009roofline} for various memory technologies and
slice memory bandwidth and compute throughputs.
As an example, Figure \ref{figRoofLine} shows the Roofline model for
three compute-intensive LSTMs with 
various balance of compute throughput and memory bandwidth. 
The figure shows the rooflines for two memory technologies coupled
with the peak throughput of companion compute logic.  
According to the characteristics of each application, the throughput
can be improved by either increasing computation rate or memory
bandwidth.  In Figure \ref{figRoofLine}, LSTM1 utilizes both memory
bandwidth and the compute bandwidth more effectively than LSTM0 and
LSTM2.  The closer peak points of LSTM1 to the balance points (knee of
the curve) underscores this observation.  In the same systems, LSTM2,
which requires more computation, underutilizes memory bandwidth.
Therefore, increasing computations at even lower memory bandwidth
improves throughput.  Alternatively, performance of LSTM0, which is
relatively (i.e., comparing to the other two LSTMs) memory-bound, can
be improved by increasing memory bandwidth.  

All in all, this Roofline model emphasizes that based on the characteristics of an 
specific architecture (e.g., FLOPs per Byte ratio and memory bandwidth) an application with high 
FLOPs:Byte ratio may be labeled as either {\em compute-bottlenecked}, or {\em bandwidth-
bottlenecked}. The key point here is trying not only to match the compute rate to memory bandwidth 
of the hardware, but also to imitate the requirements of an application by characteristics of the 
hardware.
For example, DNNs vary in their structure and computational requirements both
within and across applications (e.g., based on their structural
features such as number of connections, size of each layer, specific
neuron model, training features such as batch and input sizes, etc).
Structuring near data processing with slices produces memory systems
with a high degree of concurrency - many concurrent memory channels
between memory components and accompanying compute
logic. As a result, for problems with a high degree of fine-grained
parallelism, data partitioning and mapping algorithms can be used to
maximize the utilization of memory bandwidth and compute logic.  Since
these are memory  systems, partitioning and mapping are analogous to
memory allocation and data layout in memory systems and effective
implementations balance performance across slices to fully and
effectively exploit the fine grained parallelism available in data
intensive computations. Thus, performance scales with the size of
allocated data and hence memory.

\section{Memory Slices}
\begin{figure}[t]
\centering
\includegraphics[width=0.75\linewidth]{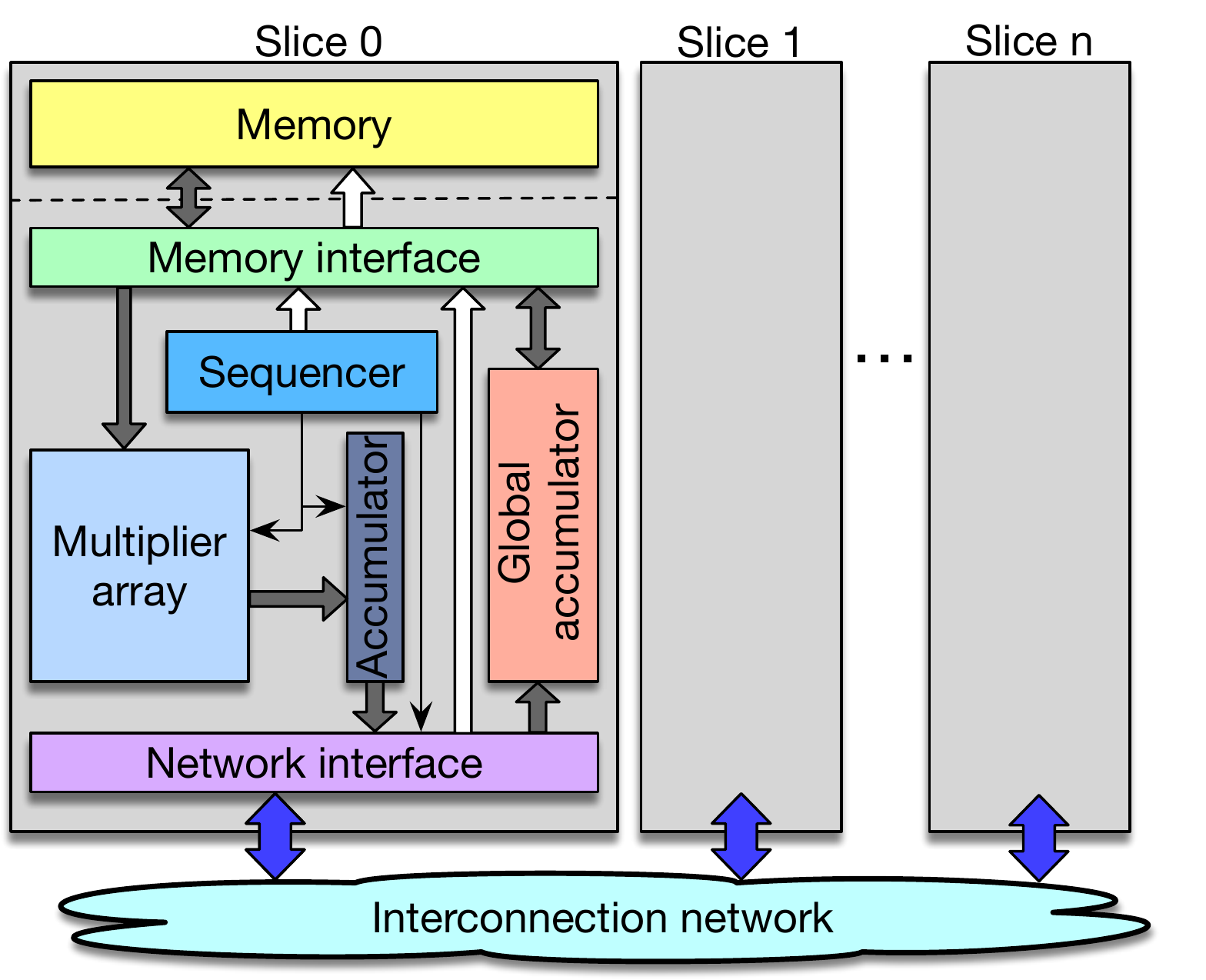}
\vspace{-10pt}
\caption{Memory slices. Gray, white, and single-line arrows indicate data, address, and control signals.}
\label{OverallArch}
\vspace{-0.2in}
\end{figure}

\noindent This section introduces the notion of memory slices and
their aggregation into larger memory systems structured around NDP. We
describe the microarchitecture of a specific type of memory slice and memory system,
used in this paper to accelerate a specific class of DNNs.


\begin{figure}[b]
\centering
\vspace{-0.23in}
\includegraphics[width=0.6\linewidth]{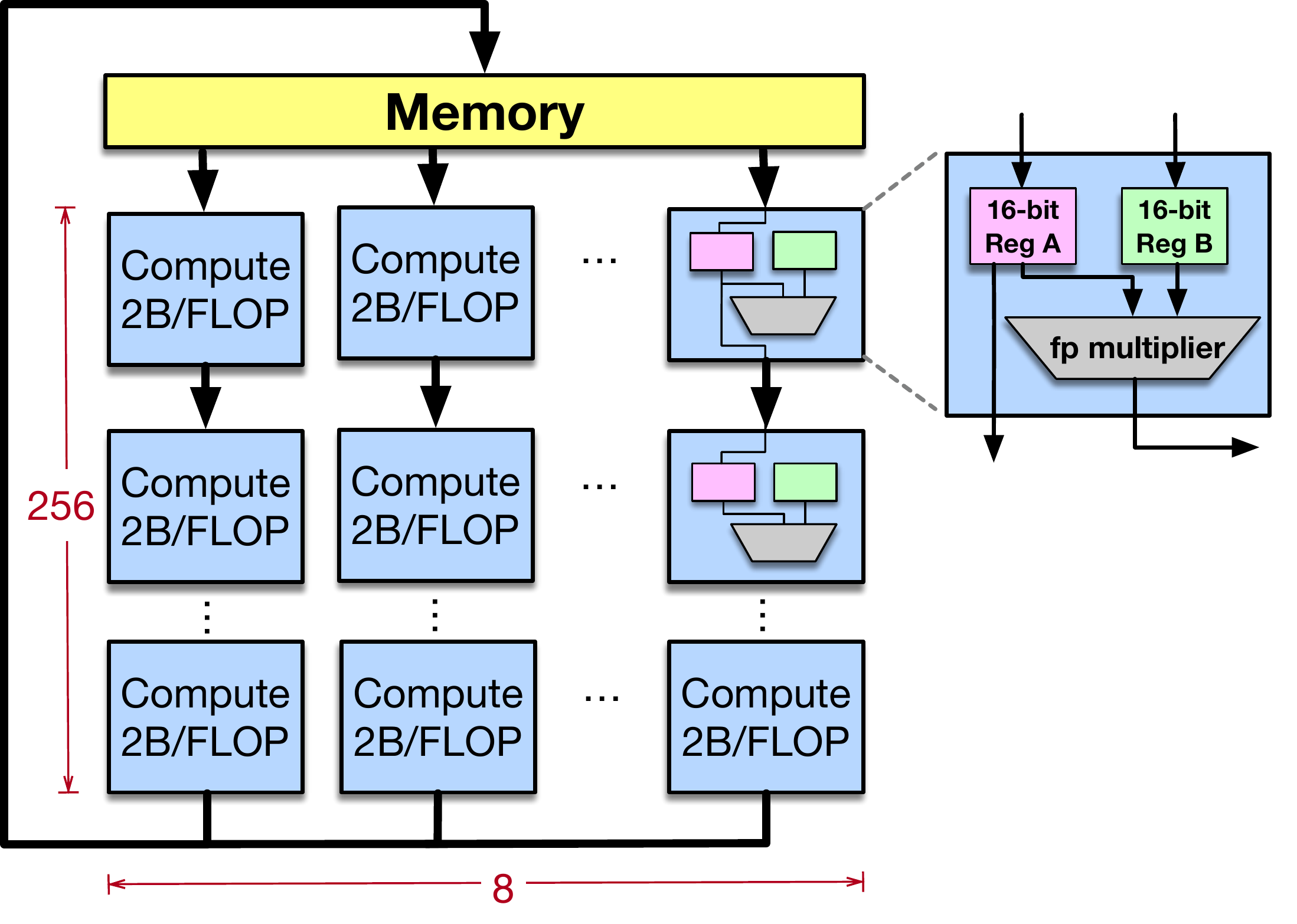}
\vspace{-15pt}
\caption{Using parallel systolic approach in a multiplier array. The Reg A of compute units include the transient values, and Reg B of them consists of the preload values. }
\label{systolic}
\end{figure}

\subsection{Slice Organization}

A slice is comprised of a bank of addressable memory, a programmable
memory interface (PMI), compute logic, and a network interface to a
wormhole switched inter-slice communication network (ICN). Depending on
the packaging, the ICN may be a network-on-chip, network-on-package,
or network-on-board. The specific slice microarchitecture used here is
illustrated in Figure \ref{OverallArch}. An important characteristic
of a slice is the ratio of the memory bandwidth to the peak compute
bandwidth. The former depends on the specific memory technology (e.g.,
DDR, HBM) while the latter depends on the compute logic (e.g., simple
cores, systolic cores, vector, etc). Computation is data driven - data
can be streamed from the memory bank to the compute logic or driven by
data packets from the network. The individual components of the specific slice architecture 
for accelerating a specific class of DNNs, are described in more detail in the
following sections.


\subsection{Inside a Slice}

\textbf{\textit{The multiplier array and the adder tree vector:}} The
dense matrix applications of interest in this paper exhibit
significant fine grained parallelism and a high ratio of compute
operations to memory accesses (see Figure~\ref{figRoofLine}). We
employ a slice microarchitecture that exploits fine-grained
parallelism and maximizes data reuse, while balancing compute
throughput and memory bandwidth. We concurrently achieve these goals
by adopting a systolic array design~\cite{gentleman1982matrix,
  kung1988vlsi} - a regular and cost efficient architecture, in which
data flows from memory, rhythmically passes through processing
elements (PEs) contributing to all partial product computations that
require that element thereby minimizing accesses to each data element.
As shown in Figure~\ref{OverallArch}, we have a 2D array of
multipliers, where each row is connected to an adder tree for
computing the sum of the products in each row. The choice of the
number of multipliers and their interconnection are tuning knobs for
adjusting the required parallelism and compute-memory balance of the
multiplier arrays and memory for an application. Figure~\ref{systolic}
shows an example of a systolic multiplier array with a 32-bit
multiplier and two source registers per processing element.  In this
configuration, eight 256-unit columns are connected in a manner
wherein at each step i) all PEs compute the product of its inputs, ii)
the row adder tree computes the sum of products, and ii) the
values in Register A are shifted vertically down to the next
PE. Register B maintains pre-loaded values (see example below). With a
2 GHz clock and 3 cycles per operation, such an organization can provide
1.28 TFLOPs/Sec with memory bandwidth of 10GB/s (this is a sample example without counting addition operations).  

\begin{figure}[t]
\centering
\includegraphics[width=1\linewidth]{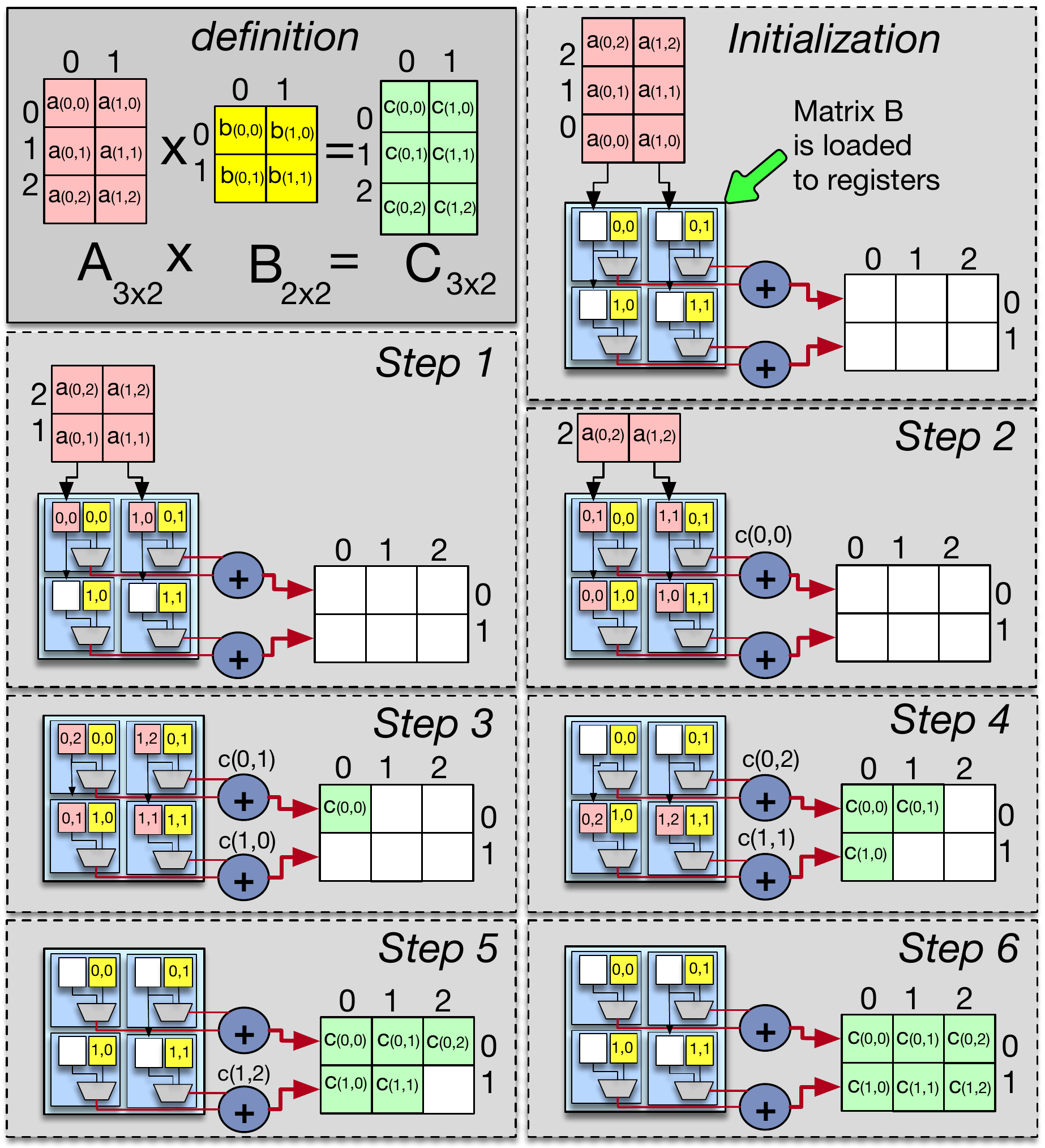}
\vspace{-25pt}
\caption{Steps of matrix-matrix multiplication.}
\label{six_step}
\vspace{-0.2in}
\end{figure}

Figure~\ref{six_step} illustrates a simple example of multiplying two tiny
matrices (A and B) and producing matrix C.  At each step all floating-point
multiplications work in parallel with the adders. Matrix B is
preloaded and column elements of matrix A are streamed through the
corresponding columns of the multiplier array. 
In the proposed implementation, we have a 256x8 array of multipliers per
slice - 8 per row. The adder actually represents an adder tree.  The multiplier latency is 3 cycles (@2GHz) and the
adder tree takes 3 cycles to add the sum of 8 products
thereby maintaining the streaming behavior. 

\textbf{\textit{The aggregation engine:}} Large problems will be
partitioned across slices necessitating aggregation of values across
slices. For example, for large matrices partitioned across
multiple slices, partial sums or inner products have to be accumulated from multiple
slices. Other operations may have to be then applied to these global
sums, for example the application of activation functions to the
computed neuron state where the state computation may have been
computed across multiple slices.

\textbf{\textit{Memory controller:}} A key components of
a slice that ease the modularity and scalability is the PMI. 
The PMI has a table that maps abstract information
such as matrix indices to physical addresses.  
In general, such abstract information include indices, tags, IDs, etc. 
for identifying parameters of applications, dependencies, subsets of data
corresponding to each step of an application, the length of vectors to
be streamed, etc.  The mapping table is populated during the
programming phase, in which the host sends configuration packets via
the ICN to setup the mapping tables of all memory slices. 
 To guaranty the correctness of operations, all the operands of one row should be 
appeared before firing one shifting of operands in registers. For dense applications with high 
locality, we assume that memory access latency will not create a bottleneck. 
In other words, in such architecture memory bandwidth defines 
performance and not the memory latency. 
As registers themselves can play the role of buffers in permitting  intermediate off-die memory latency, 
therefore, we do not use buffers. Stream-processors such as \cite{dally2003merrimac, 
mai2000smart} supports those kind streamings.

\textbf{\textit{Interconnection network:}}
During run time, the sequencer accesses the PMI table.
Then, required data is read and streamed from memory to the multiplier array. 
The network interface creates packets and sends them to the destination
slice/slices using the ICN.
The mapping algorithm helps in increasing the fraction of local data transfer (i.e., without 
accessing the ICN), locating communicated slices close to each other.
In addition, the network interface implements a
coalescing optimization where multiple data elements destined for the
same destination are encapsulated in a single packet. 
As a result, the injection rate of packets is kept low, to prevent saturations in the 256-node network. 
Note that besides these optimizations, memory slice architecture, which utilizes systolic arrays 
at NDP eliminates two typical types of traffic flow, data distribution from 
a global buffer to processing elements (PEs), and traffic from multiple PEs to the global buffer, which 
occur in most of the neural network accelerators~\cite{chen2017eyeriss, du2015shidiannao, 
jouppi2017datacenter, akopyan2015truenorth}.
The mentioned factors, impacts the traffic flowing in the ICN and in turn the efficiency of the system.

\textbf{\textit{The sequencer:}} While utilizing a cost-efficient
systolic array of multipliers is the key contribution in maintaining a 
desirable balance between compute and memory to reach high throughput, an accurate sequencer should precisely adjust and
synchronize the parallel stream inputs of  arrays to guarantee the
correct operations. In other words, all the inputs (e.g., eight
parallel streams in Figure \ref{systolic}) should arrive to
multipliers of a row at the same time, so the the vector of adder trees
can sum correct values together. The sequencer at each module
guarantees this synchronization. 
The sequencer is a programmable state machine that sequences all of
the data flow within a slice. This includes i) streaming data from the
memory to the multiplier array, ii) between the multiplier to the
adder trees, and iii) from the adder trees to the network
interface where it will be streamed back to memory or a remote slice
depending on the destination address. In doing so, the sequencer ensures
synchronization between units (e.g., all data is ready before
initiating the multiplier array). The sequencer also manages the asynchronous
interactions between the aggregation engine and the network
interface.

\section{Partitioning And Mapping}

\begin{figure}[t]
\centering
\includegraphics[width=0.9\linewidth]{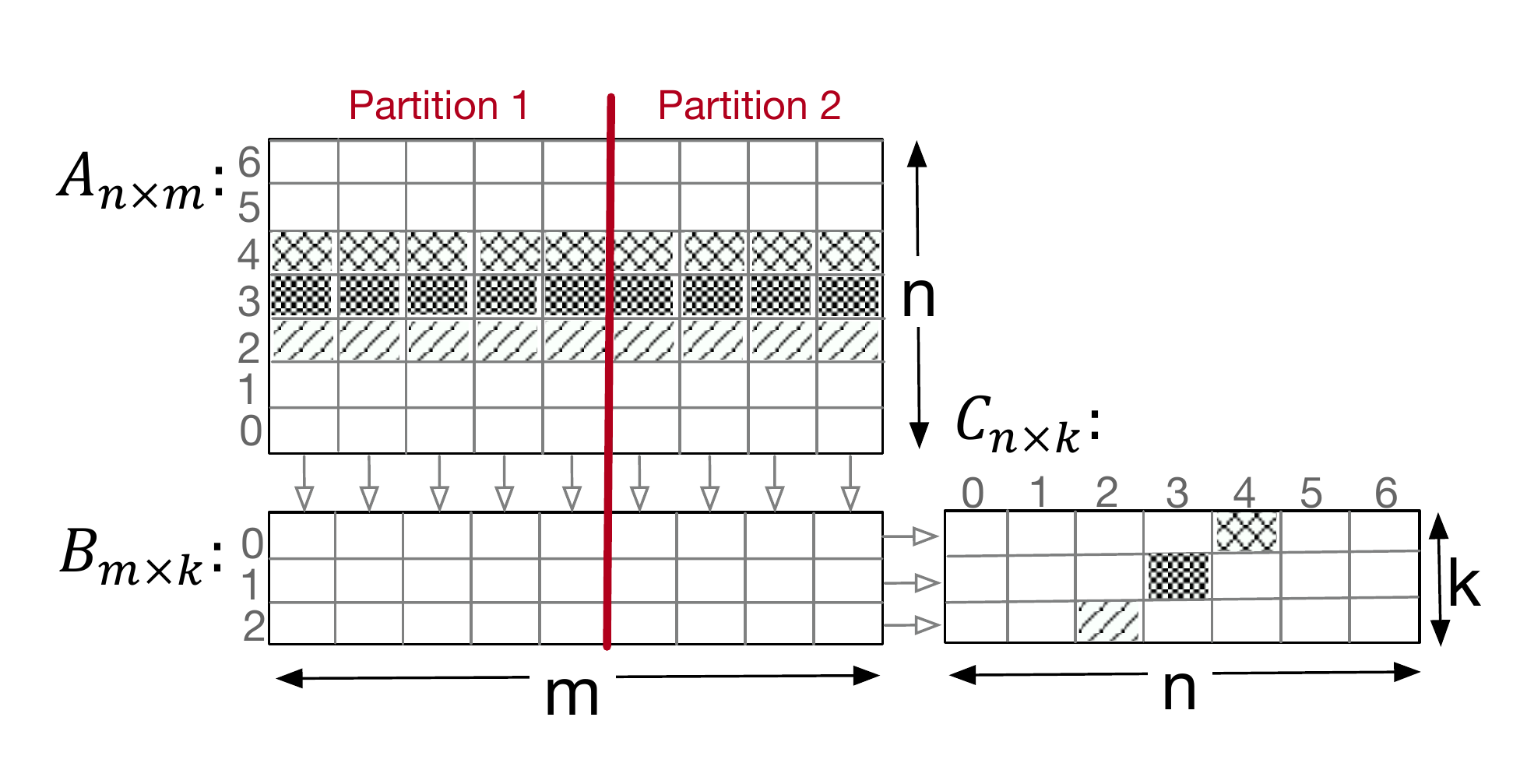}
\vspace{-20pt}
\caption{Partitioning two matrices engaged in a matrix multiplication.}
\label{matrix_multiplication}
\vspace{-0.3in}
\end{figure}

The slice-based memory system operates under the control of a host that is
envisioned to program slices via memory-mapped memory read/write
operations much in the same way that peripherals are often
configured. Considering the acceleration of training of RNNs described in
Section~5 , compilation will produce a set of
matrix-matrix operations corresponding to a batched training
formulation. The host now i) partitions and maps these matrices across
slices, ii) computes the mapping tables that will populate the PMI of
each slice (the sequencer in each slice will use this to control local
operation),iii) programs the PMI of each slice via configuration
packets,  iv) loads all of the matrices into memory, and iv)
initiate computation. The sequencers are programmed to coordinate the
execution of application phases, e.g., forward propagation,
backpropagation, etc. Slices operate in parallel and asynchronously
thus overlapping communication with computation. 
Partitioning and mapping of matrices is central to the performance
gains. The approach is described in the following section followed by
a step-by-step description of multi-slice operation that brings it all
together

\subsection{Partitioning Algorithm}

The dominant compute kernel implemented in the applications analyzed
here is matrix multiplication. The strategy for partitioning is illustrated
in~Figure \ref{matrix_multiplication}. The basic idea is to partition
the matrices across their common dimension as illustrated in the
figure, for two slices to multiply two matrices A and B. At each step, a
row partition of A can be aligned with a corresponding column
partition of B at the multipliers in a slice to perform an
element-by-element multiplication. At successive steps, i) the
products can be summed in the adder tree for that row, ii) the
partial sum can be transmitted to the the other slice to be added to
its locally computed partial sum, and iii) the row partitions are
shifted down one row. These steps are repeated producing elements of
the output matrix as shown in~Figure \ref{matrix_multiplication}. The
indices are used to determine the network address of the destination
of partial sums. 
Since the vertical streaming of elements of each column of A can be
performed in parallel, we can partition A and B into as many partitions as
necessary (even with distinct partition sizes as shown) and assign
each partition to one slice. The systolic multiplier array favors such
a partitioning scheme by preloading elements of one of the input
matrices (usually the smaller one, B in
Figure \ref{matrix_multiplication}) into one of the registers of the
multiplier array, and streaming the other matrix (e.g., A in
Figure \ref{matrix_multiplication}) via the second register.

The preceding basic approach is applied to map large matrices across a
a fixed number of slices each with a fixed array size of multipliers and adder trees
in each slice. The dimensions of the multiplier array
determines the partition size (e.g., row partition size
in~Figure \ref{matrix_multiplication}). Consequently, for very large
matrices, one slice may in fact have to process more than
one partition of the matrices with overheads for sequentially loading required data
for each partition. In addition, in some cases, larger dimensions of the
smaller matrix (e.g., $k$ in Figure \ref{matrix_multiplication}) could
preclude pre-loading into the registers of the multiplier array. In
such cases, the second matrix should be partitioned both vertically
and horizontally and be loaded iteratively to the
registers. Alternatively, for smaller matrix dimensions, multiplier
array elements will be under utilized. As long as the matrices are larger
than the dimensions of the multiplier arrays of a single slice, this
fragmented utilization of a multiplier array is limited to one slice
due to the flexibility of the partitioning and mapping approach
illustrated in~Figure \ref{matrix_multiplication}. 
In addition to partitioning matrices based on the size of 
slices, we define a mapping between the partitions and nodes (slices) which are 
connected in a network.
Note that besides partitioning, the mapping approach is central to determining the performance of 
the system because it effect the traffic of the ICN directly. 
In the examples of dense applications, studied in this paper, we heuristically map 
the partitions sequentially to the slices.


\begin{figure}[t]
\centering
\vspace{-0.24in}
\includegraphics[width=0.85\linewidth]{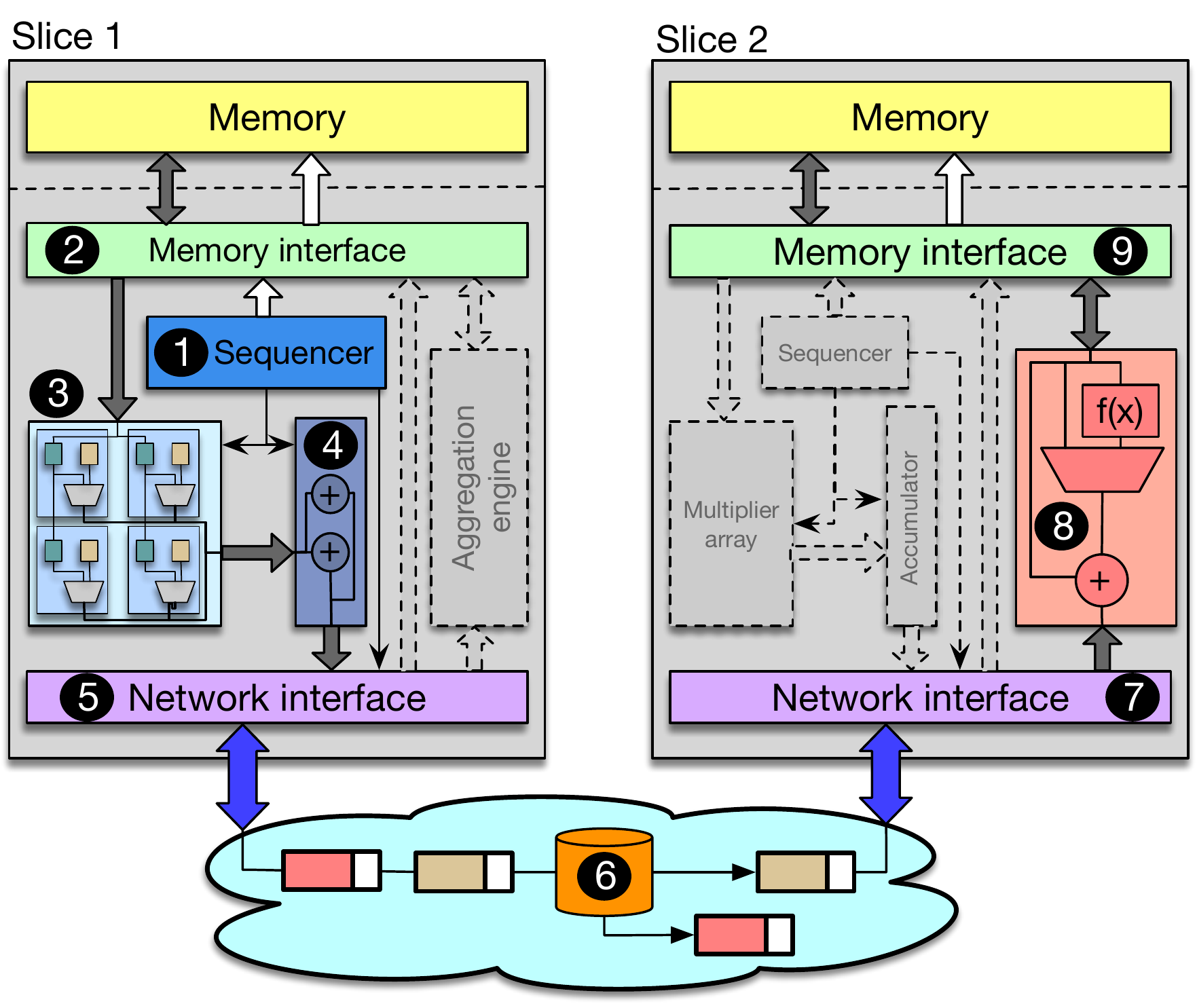}
\vspace{-15pt}
\caption{Nine steps of micro-architectural operation at two sample slices. }
\label{steps_micro}
\vspace{-0.3in}
\end{figure}

\subsection{Microarchitectural Support}

The last section provided a description of how matrices are
partitioned and mapped across slices. This section describes the
details of microarchitectural operations at each memory slice in
support of the example in Figure \ref{matrix_multiplication}. Assume
that partition 1 and 2 of Figure \ref{matrix_multiplication} are mapped
to slice 1 and 2 in Figure \ref{steps_micro}, and that the output
(matrix C) is mapped to slice 2. While both of the slices execute
concurrently, Figure \ref{matrix_multiplication} describes the nine
steps initiated in slice 1 to 2 as follows: \ding{182}\ To start a new
matrix-matrix multiplication, the sequencer pre-loads the partition of
matrix B to the systolic multiplier array and initiates streaming
elements of matrix A via the memory interface to load the next
elements to the first row of multiplier array (in general equal to
width of multiplier array). 
\ding{183}\ Once the memory interface loads new values
to registers of the first row of the multiplier, the previous values
of those registers shift one row down. \ding{184}\ All the multipliers
in the array operate in parallel, and send their results to the
adder tree unit for that row. In this example, assume slice 1 is at a
step that rows 2,3, and 4 of matrix A are at the multiplier array to
be multiplied by rows 0,1,2 of matrix B to create a portion of
elements (0,2), (1,3), and (2,4) of matrix C.
 
\ding{185}\ The local adder tree tree partially sums the results of
multipliers involved in creating these  elements of the
output. In general, partition of the output matrix may reside in the
local or remote slice. As a result, the adder tree vector, which is only aware
of the index of the output matrix (based on indices of its operands), but
not the slice number, send its results to the network interface
which has the information about the mapping of matrix partitions to
slices. In this example, the three elements are transmitted  to slice
2. \ding{186}\ The network interface packetizes data and sends them to
their destinations. In the case that the source and the destination of
a packet are the same, the network interface will route it to its
local port. Since it is already known that the elements of destination
matrix (C) are located diagonal, the network interface only sends the
abstract address (indices) of the first element in a packets and the
number of elements (e.g., (0,2) and 3). At the destination slice, the
addresses of other elements will be made by using the first
one. \ding{187}\ The ICN routes packets to their
destination. \ding{188}\ The network interface of slice 2 receives
packets and after extracting, sends them to the aggregation unit, which sums up partial sums of each element from
the slices. \ding{189}\ If the received packet includes the
last partial sum, then this unit applies other required  functions to the
results (e.g., the activation functions for
neural networks). Otherwise, the result directly updates the value of
memory. \ding{190}\ The memory controller has two
tasks here. First, it fetches the other portion of the same element from
memory, so that the aggregation engine adds the new portion to that. Second, it writes back the
values at dedicated addresses based on the mapping table. 
While the nine steps are basic steps, some complex applications may require additional 
operations. For instance, for CNNs, converting 
convolution operations to matrix-matrix multiplications duplicates some elements
of inputs. In such scenarios, the memory interface may create two or more copies of an output element at write them 
in specified elements based on the mapping table.

\section{A Dense Application Example}

\begin{figure}[t]
\centering
\vspace{-0.24in}
\includegraphics[width=0.8\linewidth]{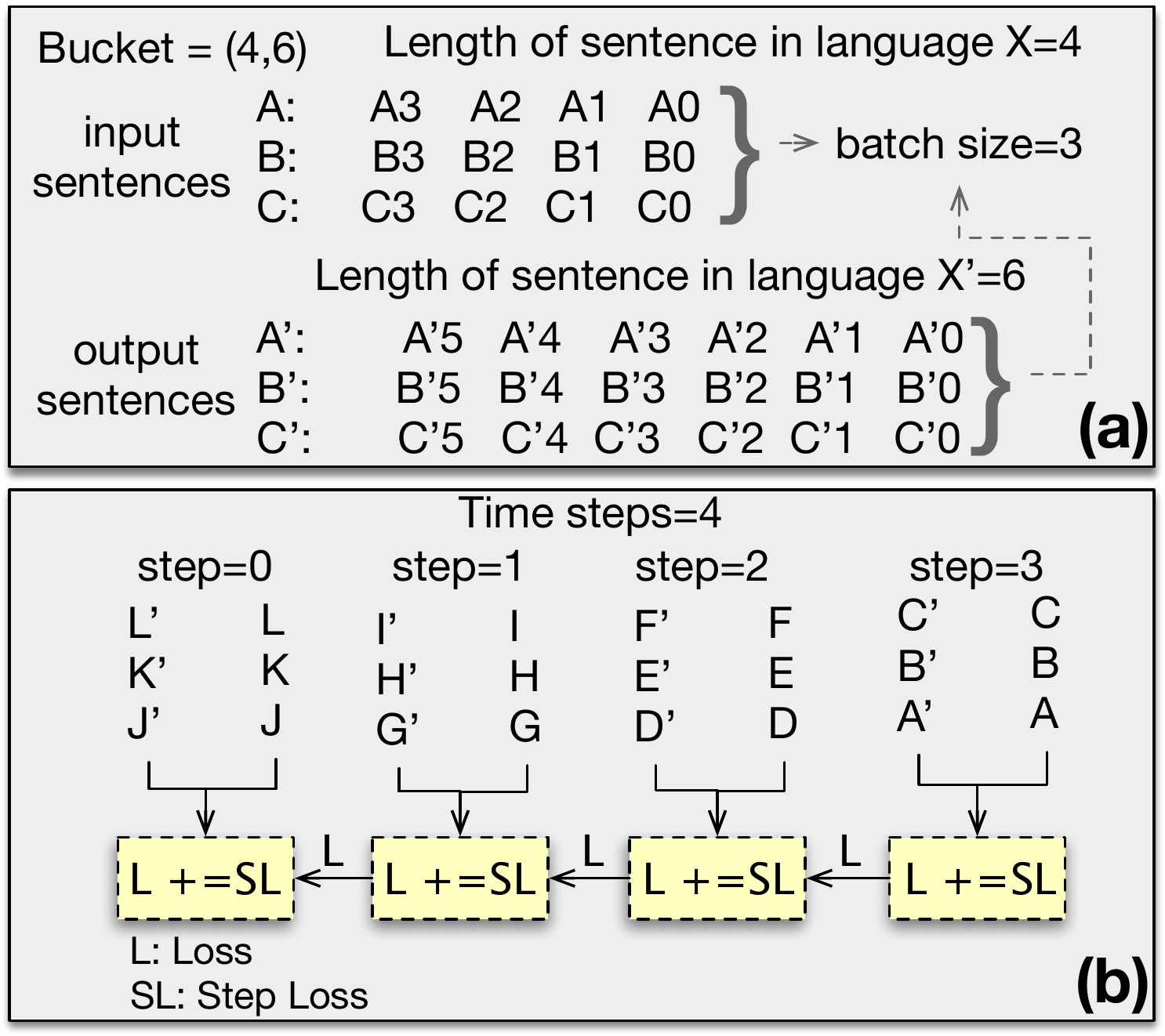}
\vspace{-10pt}
\caption{A translator example, a) A, B, and C are four-word sentences in language X, translated to A', B', and C', six-word sentences in language X'. Batch size is 3. Sentences of language X and X' are loaded to the encoders and decoders respectively. b) Four time-steps of truncated BPTT, at each a new batch is fetched from dataset. The loss are propagated through time-steps.}
\label{Translator_example_param}
\vspace{-0.2in}
\end{figure}

This section explains memory slices by using an example of RNNs and describes two major points: First, how required operations for forward and backward propagations for training RNNs are converted to matrix-matrix multiplications, favored by systolic arrays at each memory slice. And, second, how memory slices run the training and inference efficiently and correctly by maximizing parallel operations required at each step of algorithms. 
To date, for solving various ranges of sequential applications, from speech and video processing to language modeling and machine translation, several structures have been proposed, all of which are unanimous in having the feedback connection from previous elements of a sequence in their structure, while are varied in details of their structure. Predicting next word of a sentence, or translating a sentence are applications of RNNs.  In this section, we picked neural machine translation (NMT)\cite{bahdanau2014neural} Figure \ref{Translator_example_param}-a  defines the fundamental parameters of a translator. 
Since processing sentences with random sizes is almost impossible for current translators,  bucketing mechanism is used to justify sentences of various size by padding (i.e., adding special tokens to sentences for enlarging them). However, when variation in length of sentences is too much, padding all the sentences to a single size is very inefficient. Therefore, usually a group of buckets with various sizes are used In this example, we assume only one bucket pair of size (4,6).

Unlike inference, during training with teacher forcing technique\cite{toomarian1991fast}, both input and output sentences are loaded into the translator.
The size of batch is assumed as three in our example. Therefore, three sentences of a dataset will be processed in forward propagation, then the loss is calculated, and finally weights are updated and errors are back propagated.  
In RNNs error should be back propagated through time (BPTT)\cite{werbos1990backpropagation}, which suffers from vanishing gradient problem when length of sequences enlarges. To overcome this issue, truncated back propagation is proposed\cite{sutskever2013training}, which back propagates errors to only a limited number of time-steps. Figure \ref{Translator_example_param}-b shows back propagation of errors through four time-steps. In Figure \ref{Translator_example_param}-b, the basic block of the translator, which translates three four-word sentences to three six-word sentences, is unrolled to four time-steps. At each time-step, the translator block processes a new batch by fetching three sentences of input and three corresponding sentences from output; and updates the loss by adding its own loss (step loss) to the back-propagated loss. All the mentioned dependencies will be defined at the table of PMI, so that the sequencer can keep the track of them. 

\begin{figure}[t]
\centering
\includegraphics[width=1\linewidth]{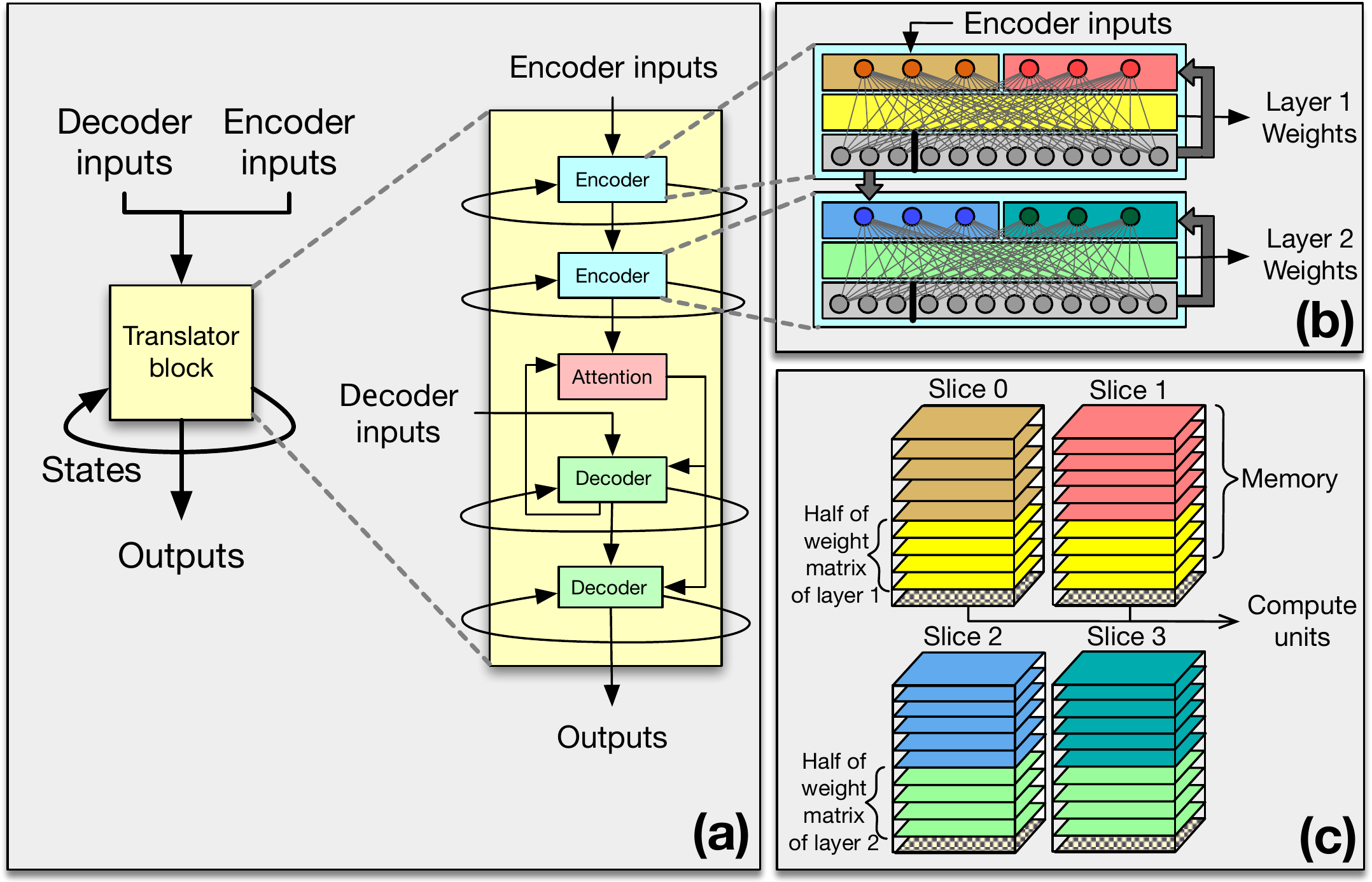}
\vspace{-15pt}
\caption{a) Translator layers, including two decoders and two encoders (each an LSTM), and a feedforward attention layer. b) Nodes and connections of an LSTM. c) Partitioning two layers across four slices. The weight matrix of each layer is split across two slices.}
\label{partitioning_layers}
\vspace{-0.2in}
\end{figure}

A translator block can consist of a stack of RNN cells for increasing the accuracy. Figure \ref{partitioning_layers}-a illustrates the block diagram of one translator block including five layers, four of which are LSTMs and one(the attention) is a feed-forward network. This example is a simplified version of attention-based translators, a mature version of encoder-decoder-based translators\cite{cho2014learning}. First, encoders process input sentence of first language. The input of an LSTM is an input word and a previous state, represented in one vector, which is multiplied by the weight matrix. 
The output of last decoder layer goes to the attention layer, which is added to encoder-decoder-based translators to enable translating long sentences\cite{bahdanau2014neural}. Similar to encoders, each decoder cell is an LSTM.

\begin{figure}[b]
\centering
\vspace{-0.24in}
\includegraphics[width=1\linewidth]{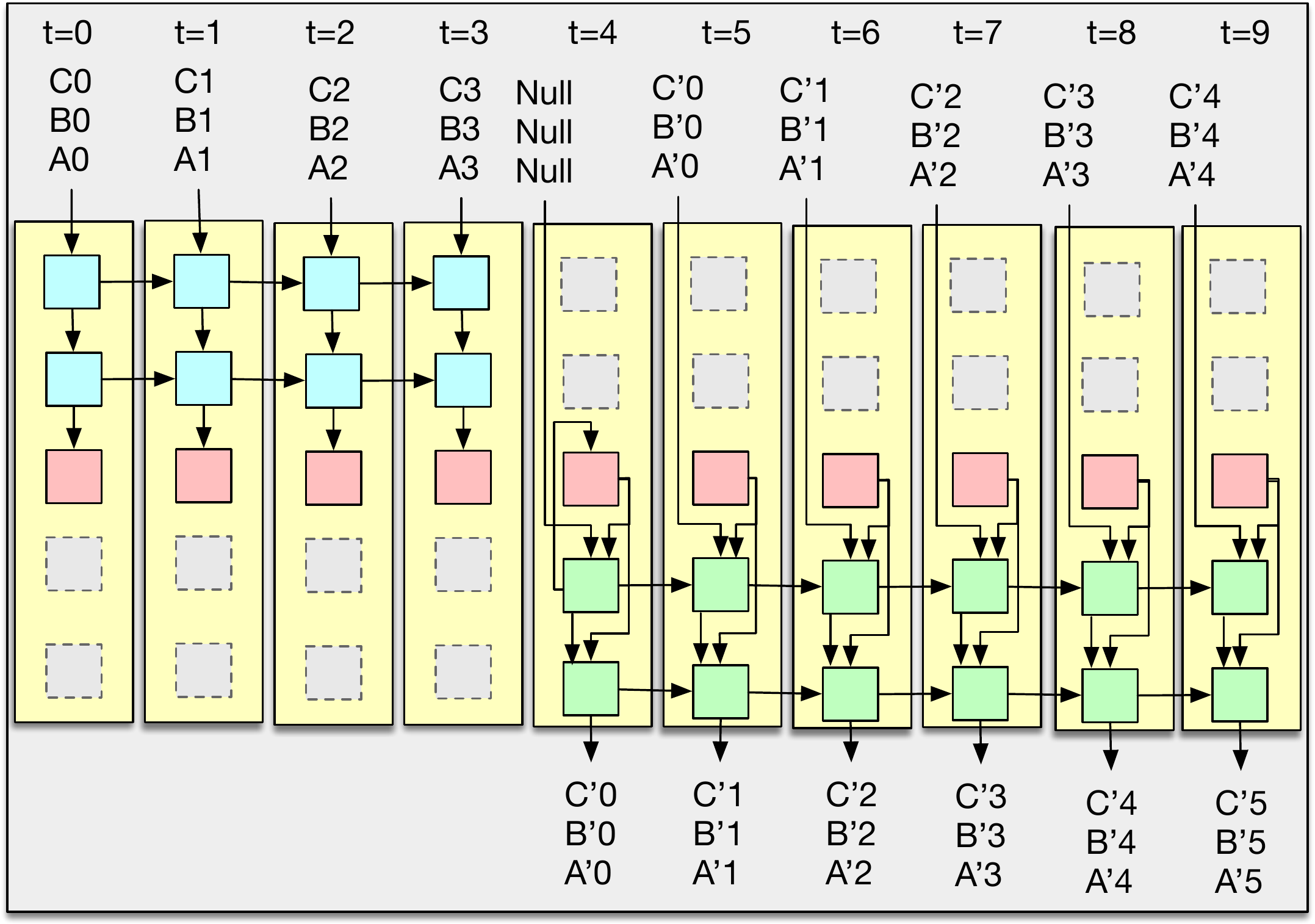}
\vspace{-20pt}
\caption{Ten times unrolled translator block at time-step=3 of Figure \ref{Translator_example_param}-b. The colored boxes are activated and dashed gray boxes are inactive at a micro-step.} 
\label{unrolled}
\end{figure}

\begin{figure*}[t]
 \center
  \includegraphics[width=\textwidth]{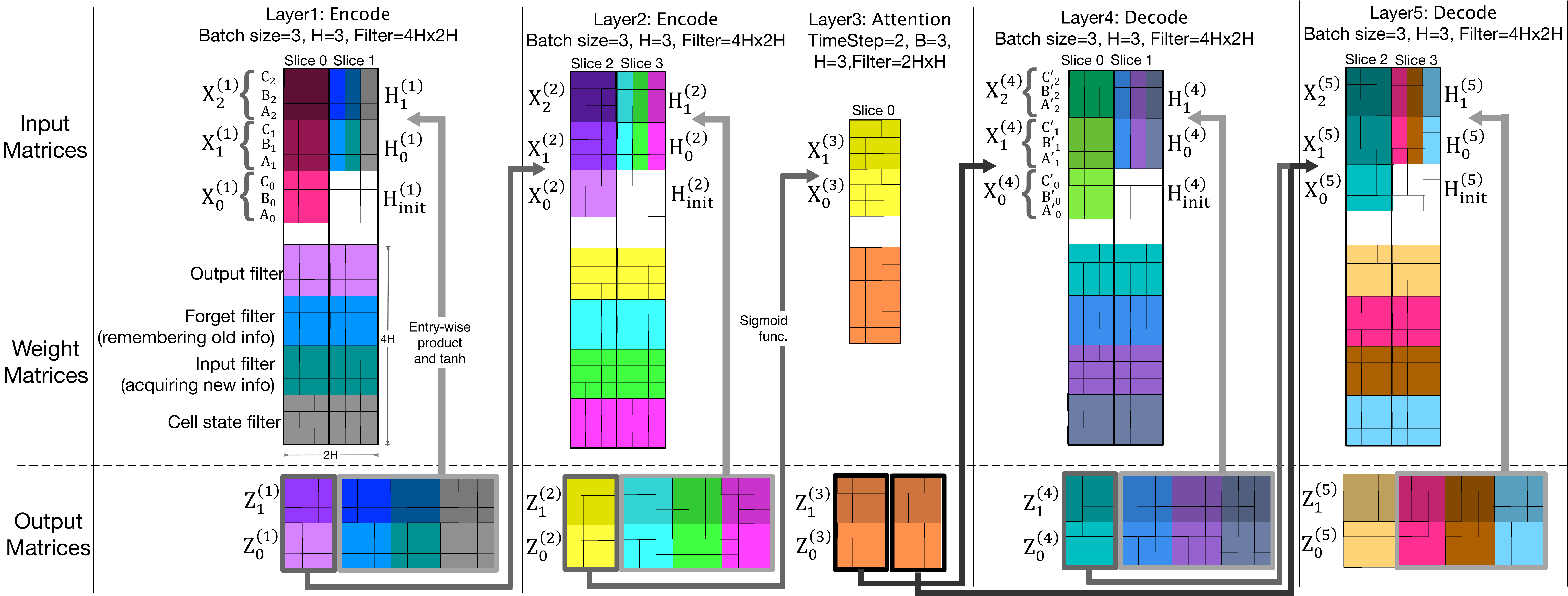}
  \vspace{-20pt}
  \caption{Matrices of translator layers, mapped to four slices for forward propagation multiplications. $X_0^{(1)}$ includes the first words (i.e., $A_0, B_0, and C_0$) and $X_1^{(1)}$ includes the second words (i.e., $A_1, B_1, and C_1$) of three sentences (i.e., A, B, C). $X_0^{(1)}$ and $X_1^{(1)}$ are inputs of first layer of translator (the first decoder). $X_0^{(3)}$ and $X_1^{(3)}$ are inputs of third layer of translator (the first encoder) corresponding to $A'$, $B'$, and $C'$ sentences.}
  \label{RNN_5layer_Exp_FW}
  \vspace{-15pt}
\end{figure*}

Figure \ref{partitioning_layers}-b displays the nodes and connections of the neural network inside two LSTM layers (the network of decoders are similar). The arrows show that the output of each layer contributes in creating the half input of next layer as well as the state part for next element in the sequence by the same layer. Figure \ref{partitioning_layers}-c shows a high abstraction of partitioning across four slices. In this example, weight and input matrices of two layers are partitioned similar to those in Figure\ref{matrix_multiplication}. As Figure \ref{partitioning_layers}-c shows, layer 1 is mapped to slices 0 and 1, and layer 2 is mapped to slices 2 and 3. As a result, in this simple scheme, slice 1 writes its results on its own memory section and on slice 2. Note that since operations of decoders will not be started until encoders are finished, the resources of encoders can be mapped to the same four slices.

\subsection{Training Neural Networks by Memory Slices}

This section explains how memory slices ease training, for proceeding which, each time-step shown in Figure \ref{Translator_example_param}-b is unrolled to several micro-steps, at each a word of sentence is processed. Figure \ref{unrolled} shows micro-steps for translating a batch of three sentences at time-step=3 of Figure \ref{Translator_example_param}-b. Since decoders cannot start their operations until encoders are done, the number of micro-steps required for translating a four-word sentence to a six-word sentence is ten. This section continue by explaining the flow of operations for processing sentences A, B, C, and A', B', C', performed at memory slices. In the next section, to simplify the example, we assume that all the sentences in both languages have two words, which cause four micro-steps per time-step.

\subsubsection{Forward Propagation}
Figure \ref{RNN_5layer_Exp_FW} demonstrates inputs, weights, and output matrices of five-layer translator mapped to four slices when all the micro-steps of time-step=3 are completed. The upper and lower indices indicate the layer number and order of a word in a sentence. For instance, $X_0^{(1)}$ is the input of layer one (i.e., the first encoder). 
Each word is represented by a vector of length three, the so-called hidden unit size (H=3). Note that a layer, called embedding transfers one-hot words into words of hidden unit size\cite{levy2014neural}. As Figure \ref{RNN_5layer_Exp_FW} shows, in addition to $X_0^{(1)}$, $H_{init}^{(1)}$ participates in making the input matrix of the first layer. $H^{(1)}$ is the feedback state of layer one from previous elements. Figure \ref{RNN_5layer_Exp_FW} also shows that weights of LSTM layers (i.e., encoders and decoders) have size of $4H \times 2H$, made by concatenating four group of filters for creating output, acquiring information from input, forgetting history, and updating the state\cite{hochreiter1997long}. At each micro-step, an LSTM generates its output and next state by multiplying input matrix of size $H \times 2H$ to its weight matrix of size $2H \times 4H$, which results a $H \times 4H$ matrix such as $Z_0^{(1)}$ for the first micro-step. Then, the first part of this matrix, which is the result of multiplying $X_0^{(1)}$ by the output weight, is activated and it creates the input of the next layer, $X_0^{(2)}$. The remaining of $Z_0^{(1)}$ matrix participates in creating $H_0^{(1)}$, the state for processing the next element in the sequence. The same procedure repeats for layers 2, 4, and 5. However, the procedure of layer 3, the attention layer differs, which receives outputs of both group of elements (i.e., $Z_0^{(2)}$ and $Z_1^{(2)}$ that create $X_0^{(3)}$ and $X_1^{(3)}$ ). As a result, before the attention, three micro-steps should be done: 1)$X_0^{(1)} \times W^{(1)}$ , 2)$X_1^{(1)} \times W^{(1)}$ and $X_0^{(2)} \times W^{(2)}$, and 3)$X_1^{(2)} \times W^{(2)}$ . Then the attention stars working by multiplying its input matrix to its weight matrix of size $2H \times H$.

Accordingly the following are some of the considerations for training. First, the memory space for a part of input matrices (those for feedback states) that are not ready during programming phase, kept reserved in PMI table, so that the sequencer knows where to write the result. Second, as matrix multiplications for next element of a sequence for RNNs cannot be proceed before the previous one are finished, to fill the gap between processing two elements, all the same-index elements of all sequences at one batch are arranged to be streamed together during programming, to better utilize the compute unit reuse. Therefore, batch size matters for gaining throughput. Third, during programming phase, the order of loading weight matrices to registers of the systolic array defines the order of producing input of next layer and the next state of current layer (more importantly when the length of weight matrix is larger than length of multiplier array), it is considered to improve performance. 


\begin{figure}[t]
\centering
\vspace{-0.1in}
\includegraphics[width=0.8\linewidth]{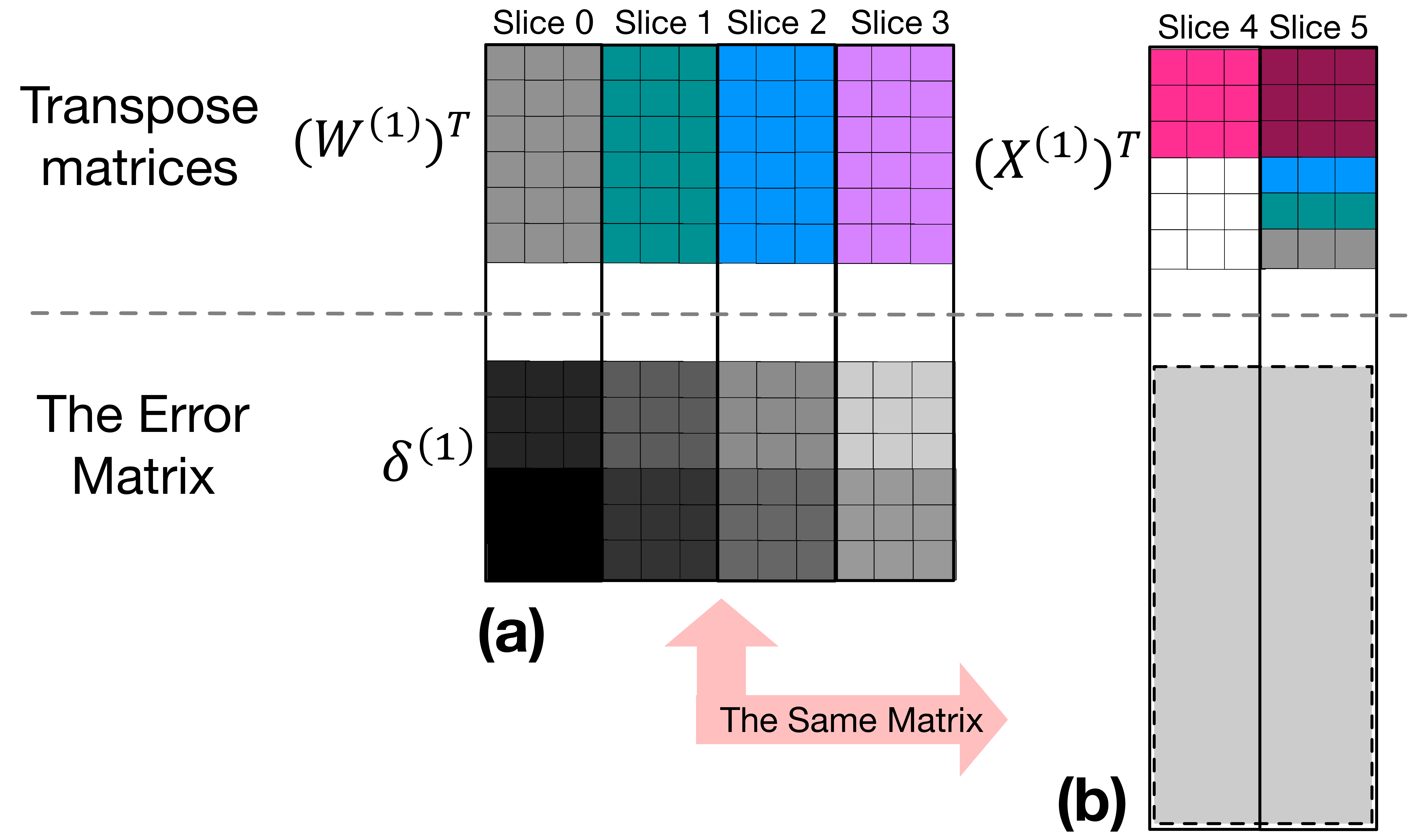}
\vspace{-15pt}
\caption{Transpose and error matrices of layer 1 of the translator, mapped to six slices for two matrix multiplication: a) for computing the error for next layer (here the error to be propagated to the next time-step), and b) for updating the weight of current layer.}
\label{RNN_5layer_Exp_BW}
\vspace{-0.2in}
\end{figure}

\subsubsection{Error Back Propagation and Weight Update}

Once all the micro-steps of a time-step are done, the back-propagating error of a layer ($\delta$), defined as scalar multiplication of activation function differentiation of layer output to matrix-multiplication of error from previous layer to matrix transpose of weight, is computed. 
In addition to error computation, weights of each layer is updated as $W=W \pm \eta (\partial J/\partial W)$,  where $\eta$ is the learning rate. 
Therefore, a pair of matrix-matrix multiplications (for calculating error and updating weight of current layer) should be performed. 
At step five (Figure \ref{RNN_5layer_Exp_BW}), the calculated error will be propagated to the next time-step. 
However, based on hardware constrains, memory slices can map them to less number of slices. 
For training, the mapping of matrices, mapped to slices in a forward-friendly manner (i.e., in a way that they are streamed into the multiplier arrays for forward propagation) should be changed to become proper for back propagation. 
To handle two mappings of same matrices for forward and back propagation, two options is possible. First, remapping matrices of each layer right after the operations for that layer in forward propagation is over. This solution uses bandwidth of memory and ICN. The second solution is mapping two copies of each matrix during programming phase. As long as the memory space is not limited, the second approach is preferred.


\section{Experimental Setup}

\noindent 
For our evaluation, we configure memory slices based on models of 3D
DRAM technologies - the Hybrid Memory Cube
(HMC)~\cite{hybrid2013hybrid} and High Bandwidth Memory
(HBM)~\cite{standard2013high}. Other packaging options are also
feasible (e.g., DDR, 2.5D, etc.) but are not evaluated here as we do
not expect the main insights to differ although specific design
choices should be affected.
We evaluate a memory system constructed of slices using an in-house
cycle-level simulator. For the RNNs and hybrid networks we simulate
the forward and backpropagation. In addition, to estimate the power
consumption integrate into the system the Kitfox1.1
library\cite{song2015kitfox} at the 16nm technology node. Kitfox is a
library that incorporates a number of public domain power models
including CACTI and McPAT - we use the McPAT model\cite{li2009mcpat}
for the power modeling of compute units and estimates of access energy
per bit for the DRAM based on 6pJ/bit for HBM~\cite{ o2014highlights} and 3.7pJ/bit for HMC~\cite{jeddeloh2012hybrid}. Table \ref{ICN} lists the
details of the ICN and the configuration of a single slice which is comprised
of an $8\times256$ 16-bit multiplier array.  Each slice is assumed to
be on the order of 1Gbytes in size and is large enough to host
the partitioned data sets considered in these workloads. 

We define eight system configurations for evaluation as listed in
Table\ref{Cfg}, modeling two versions of HMC-style and HBM-style
memories.  The first four configurations employ the basic compute unit
(i.e., that in Table\ref{ICN}) and are referred to as the baseline
configurations. The other four configurations, which we refer to as
{\em balanced} configurations increase the compute throughput with the
addition of up to 2.5x compute units/slice. This rebalances the ratio
of memory bandwidth to compute bandwidth in a slice to be closer to
the knee of the Roofline models of the evaluated configurations that
employ these memory technologies and host these applications.  Each
slice nominally represents one HMC-style vault, or HBM-style channel.
The memory system configurations that use balanced slice configurations
have the same peak TFLOPs/S as the memory system configurations that
use the baseline slice configurations, but with a smaller number of
slices. We evaluate training performance of four LSTMs and four CNNs. The proposed partitioning and
mapping algorithm optimizes each layer separately  (each set of
matrices). Consequently, it is straightforward to apply to hybrid
networks with any number of convolutional layers combined with any
number of recurrent layers.  However, in this paper, we evaluate the
CNN and RNN layers separately using well-known workloads to enable
comparisons with previous studies.  The four LSTMs are multi-layer
translators with varied parameters listed in Table \ref{LSTM}, trained
on the WMT'15 dataset\cite{WMT} and bucket sizes of (5, 10), (10, 15),
(20, 25), (40, 50). LSTM0 has parameters similar to Google NMT
(GNMT)\cite{wu2016google}. Four CNNs are
AlexNet\cite{krizhevsky2012imagenet}, VGG16\cite{simonyan2014very},
ResNet152\cite{he2016deep}, and
InceptionV3\cite{szegedy2016rethinking}, trained on ImageNet
dataset\cite{deng2009imagenet} and batch size of 128.

\begin{table}[t]
\centering
\caption{Parameters of ICN and the basic compute unit.}
\includegraphics[width=0.7\linewidth]{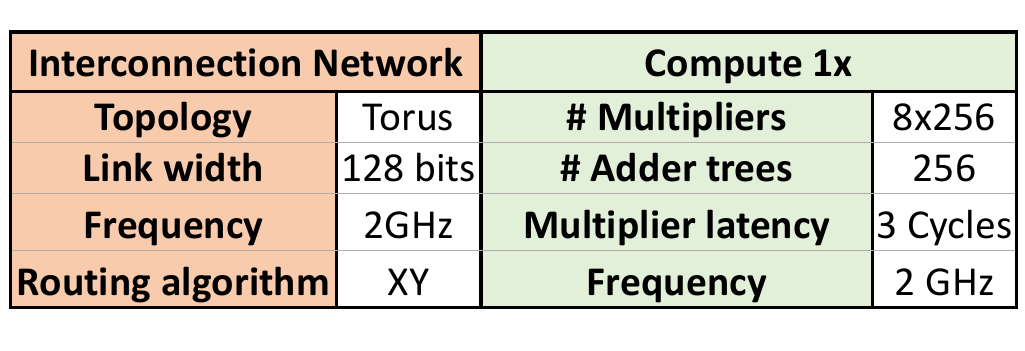}
\label{ICN}
\end{table}

\begin{table}[t]
\centering
\vspace{-0.15in}
\caption{Configuration used for evaluations.}
\includegraphics[width=0.8\linewidth]{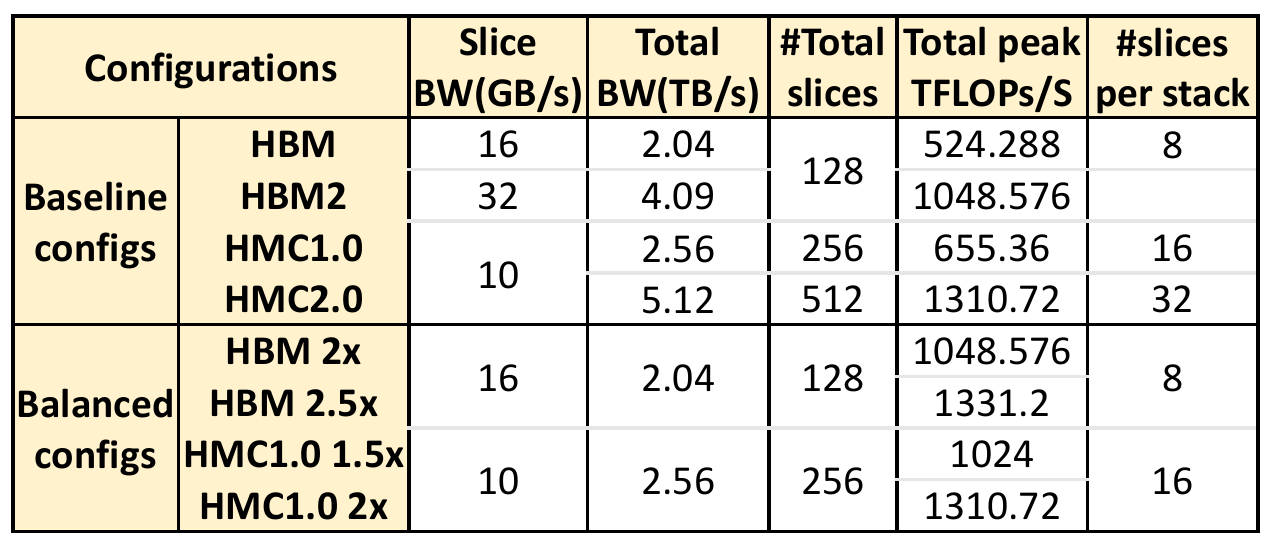}
\label{Cfg}
\vspace{-22pt}
\end{table}

\begin{table}[b]
\centering
\vspace{-0.3in}
\caption{Parameters of LSTM networks.}
\includegraphics[width=0.7\linewidth]{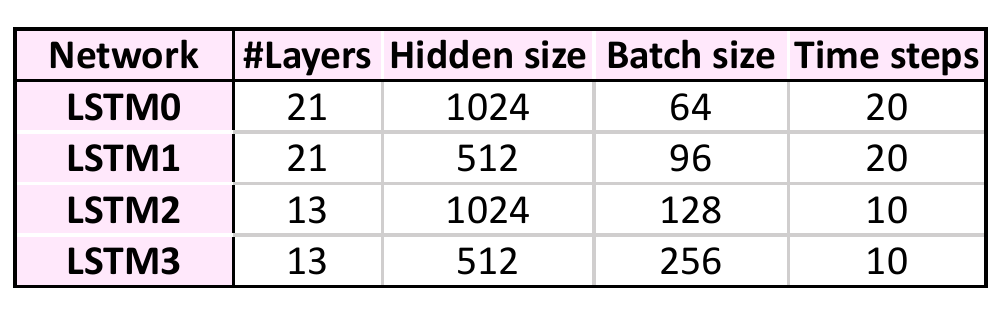}
\label{LSTM}
\end{table}

\section{Performance Evaluation}
\noindent

\begin{figure}[t]
\centering
\includegraphics[width=1\linewidth]{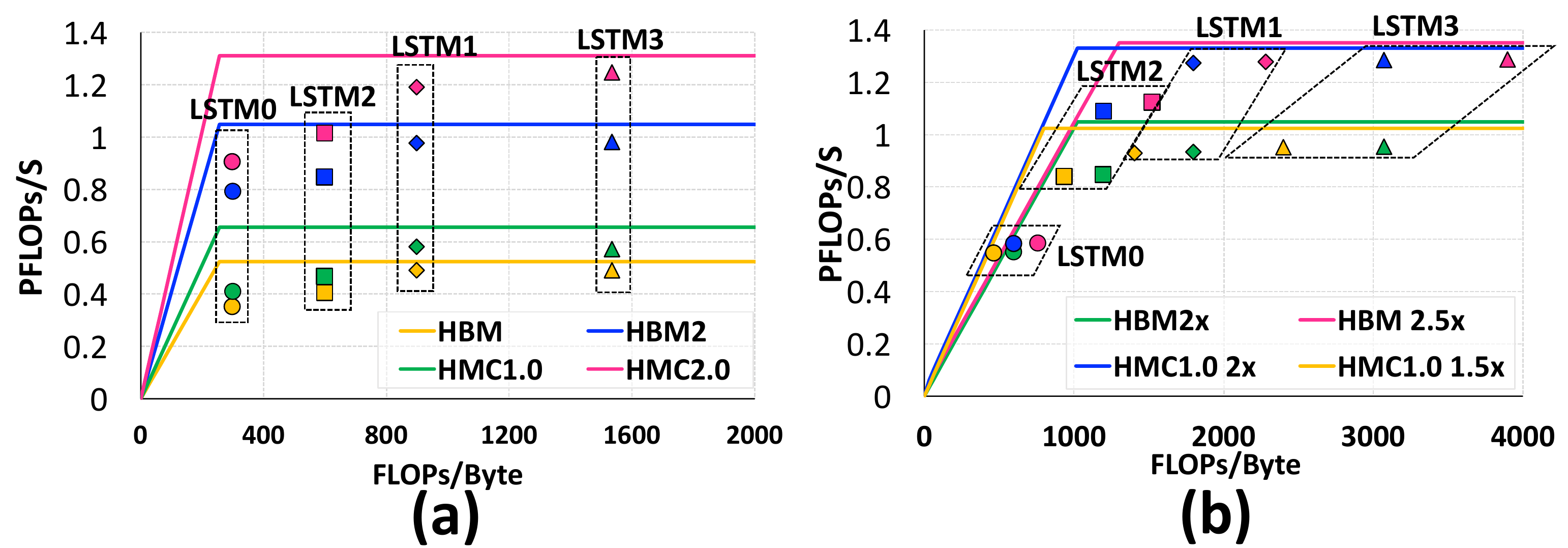}
\vspace{-20pt}
\caption{Comparing achieved and peak throughput on (a) Baseline, and
  (b) Balanced configurations} 
\label{Throughput}
\vspace{-0.05in}
\end{figure}

\begin{figure}[t]
\centering
\includegraphics[width=1\linewidth]{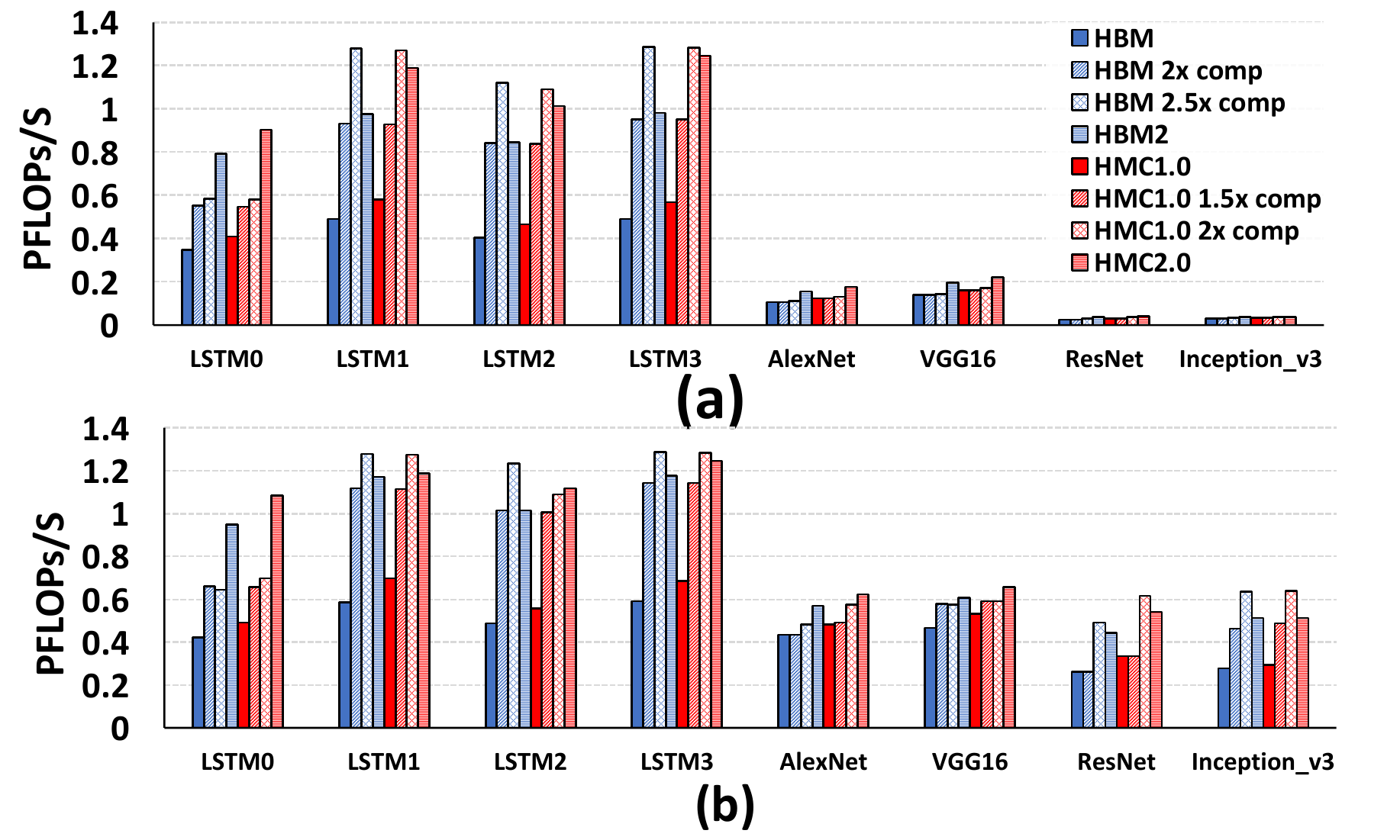}
\vspace{-20pt}
\caption{Throughput of (a) training, and (b) inference of eight
  applications on eight configurations: } 
\label{GFLOPS_train_inf}
\vspace{-0.2in}
\end{figure}

In this study we seek to learn specific lessons from executing
training of RNNs in the memory system, and the consequences for
throughput, scalability, and energy efficiency. Note that we are
studying memory {\em systems} and study specific choices of the
balance between memory bandwidth and compute bandwidth in a slice. When
we normalize the compute throughput of the memory system, different
choices will produce different number of slices to construct the
memory system. We assume the memory capacity of a slice can be
varied to keep also the memory system size constant.

\subsection{Throughput}
We locate the throughput (measured from simulation) of LSTMs on the
RoofLine model in Figure \ref{figRoofLine} - note each Roofline has a
number of slices and characteristics shown in Table 2.  Figure
\ref{Throughput}-a illustrates the results for baseline
configurations, all of which have the same FLOPs/Byte since the
configuration of each slice is fixed, the FLOPs/Byte is determined
by the operation the dimensions of the multiplier arrays and adder
trees. However the memory bandwidth differs across each
configuration. In Figure \ref{Throughput}-a we see the LSTMs are
clearly compute-bound. They do not achieve the peak throughput due to
i) dependencies in the models, ii) non-uniform sizes of layers, iii)
dimensions of matrices of a model are not perfectly matched to the
multiplier array size in the slices, and iv) performance of the ICN,
which connects the slices.  Utilization can be moderated by the
partitioning scheme or by modifying effective parameters. For example,
increasing the batch size can reduce the gap with peak throughput,
made as a result of sequential dependencies during forward
propagation. The larger batch size of LSTM3 is the main reason for its
higher throughput compared to that of LSTM2 in Figure \ref{Throughput}.

The throughput gains in Figure \ref{Throughput}-a come at some
inefficiency in bandwidth utilization.  The balanced configurations,
which combine lower memory bandwidths (i.e., HMC1.0, and HBM) with
high computation rates (again note the number of slices are different
as defined in Table 2) have different rate of data reuse (due to
larger multiplier arrays) and therefore reduce the of accesses per a
byte. As a result the working points in Figure \ref{Throughput}-b are
distributed in x-axis closer to the knee of the curve - the balance
point. A comparison between Figure \ref{Throughput}-a and b
illustrates two insights: First, those applications that were limited
by the low compute rate (e.g., yellow and green points of LSTM 1,2,
and 3 in Figure \ref{Throughput}-a) reach higher throughput by using
lower bandwidth memories but higher compute rate with consequent power
advantages. Second, even though LSTM1, 2 and 3 may not gain
significantly higher throughput by using balanced configurations, they
are located closer to the balance point and improve memory
utilization. For example, LSTM1, which reaches 1.2 PFLOPs/Sec by using
512 slices of HMC2.0, where it is 644 FLOPs/Byte from balance point,
can reach the same throughput by using only 128 slices of HMC1.0 2x,
where it is 144 FLOPs/Byte closer to the balanced point. We argue that
such insights are very useful in determining how to construct memory
systems where performance (perhaps most efficiently) scales with
memory size for an application domain.

\begin{figure}[b]
\centering
\vspace{-0.24in}
\includegraphics[width=0.8\linewidth]{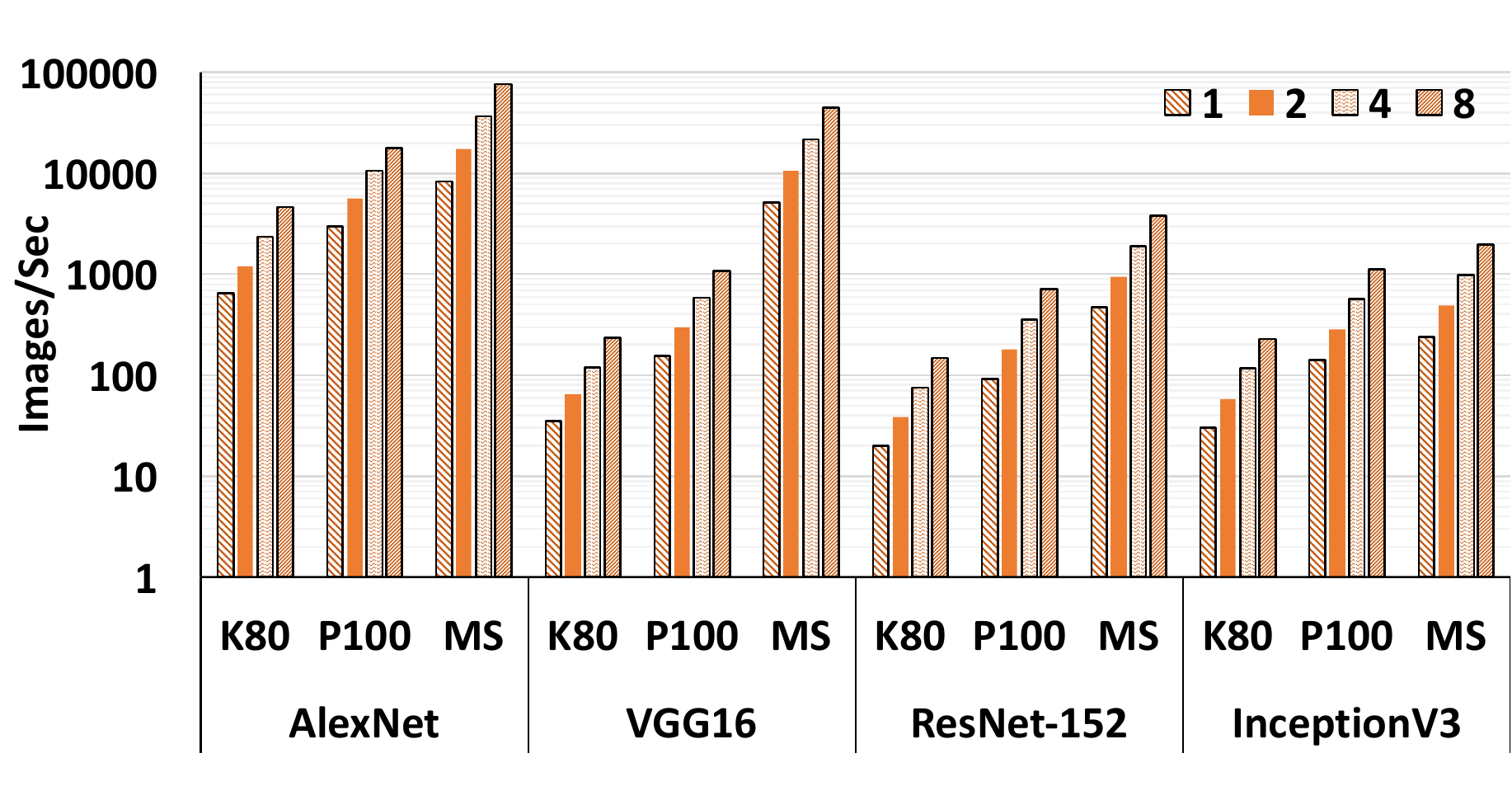}
\vspace{-20pt}
\caption{Performance of training CNNs (Images/Sec)}
\label{GPU}
\end{figure}

\begin{figure}[t]
\centering
\includegraphics[width=0.7\linewidth]{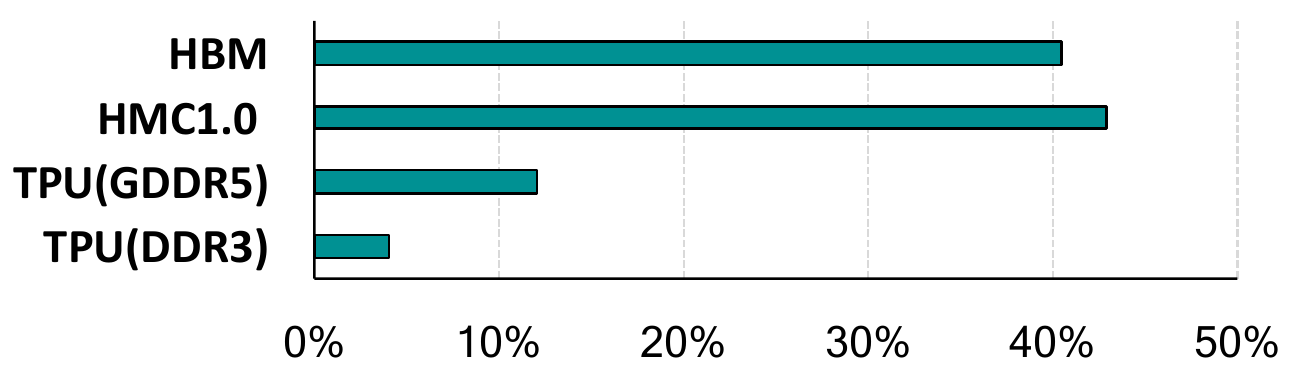}
\vspace{-10pt}
\caption{Ratio of LSTM0 throughput to peak throughput(TOPs/S) }
\label{TPU}
\vspace{-0.2in}
\end{figure}

Figure \ref{GFLOPS_train_inf} provides PFLOPs/S of all workloads for
both training and inference.  
As it shows, using slices with 2X the computation rate of the baseline increases 
performance but it does not necessarily achieve 2X system throughput. However, it is not 
orthogonal with this fact that the target application of the case study (matrix multiplication)
is heavily compute bound.
For inference we use batch processing
that is an approach for inference in server
applications\cite{jouppi2017datacenter, canziani2016analysis}. 
However, the slow error back propagation causes lower performance of
training compared to inference. Figure \ref{GFLOPS_train_inf} also
shows that the throughput of training LSTMs is higher than that of
training CNNs. There are two reasons for such behavior. The first is
uniformity in the structure of LSTMs, which helps create higher
utilization in all columns of the multiplier arrays. The second is
higher opportunity of concurrency in multi-layer LSTMs. As mentioned
earlier, once one layer of an LSTM is processing an input of a
temporal sequence, the other layers can continue processing other
elements of the sequence.

To better evaluate the throughput of memory
slice systems, we compare the number of images that can be trained per second with
Tesla\textsuperscript{\textregistered} P100 and K80 GPUs reported by
the TensorFlow benchmarks\cite{TensorFlow_Bench}. 
Since in this part we compare simulation results with real system,
 we try to do so with as mush fairness as possible.
For this reason, we have very similar
peak throughput for each pair of this comparison - four memory slices
with HMC1.0 2x has the same peak throughput as one P100 GPU.  In
addition, here we changed our batch sizes to match with those reported
in~\cite{TensorFlow_Bench}. As Figure \ref{GPU} shows, memory slices
perform similar to P100 for training InceptionV3. However, VGG16 can
gain up to 41x higher throughput when it is trained using memory
slice system. 
We also compare the throughput result of LSTM0 inference of
HBM- and HMC1.0-based memory slices with that of the TPU, using DDR3
and GDDR5. Seeking a fair comparison we choose a number of slices to
have close to the peak throughput of the TPU and we normalized results
to the peak throughput of each device. As Figure \ref{TPU} shows,
memory slices offer better performance the main reason is that the
throughput of TPU(more specifically for DDR3-based TPU) is limited by
the memory bandwidth\cite{jouppi2017datacenter}.

\begin{figure}[t]
\centering
\vspace{0.1in}
\includegraphics[width=0.8\linewidth]{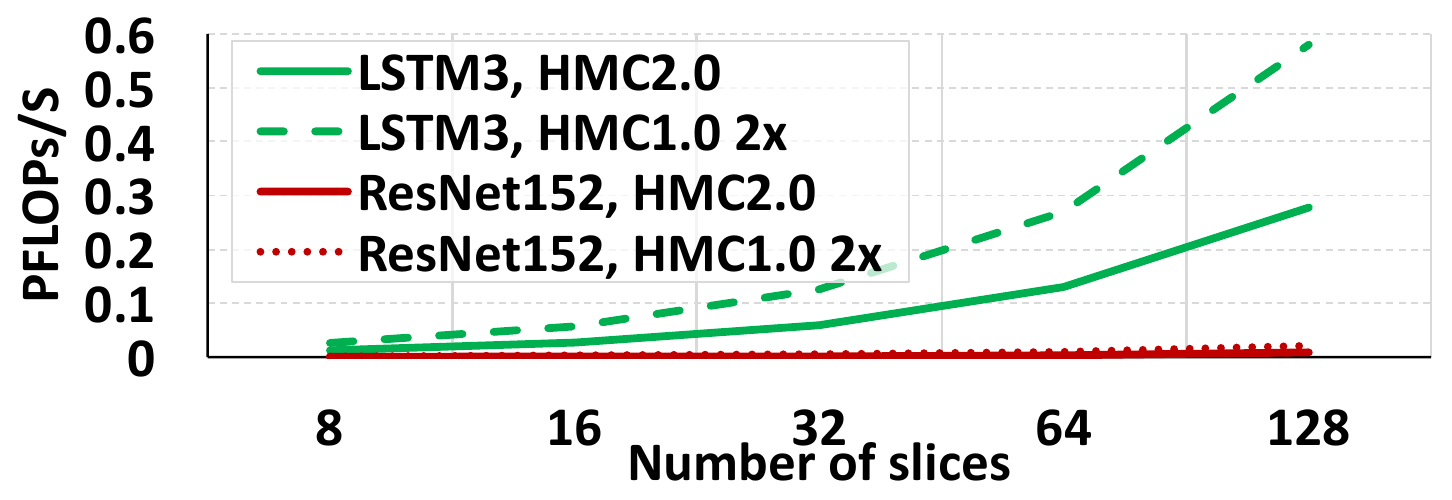}
\vspace{-15pt}
\caption{The effect of balanced architecture on scaling performance
  (GFLOPs/S), for two sample applications} 
\label{scaling}
\vspace{-0.2in}
\end{figure}

\begin{figure}[t]
\centering
\includegraphics[width=0.85\linewidth]{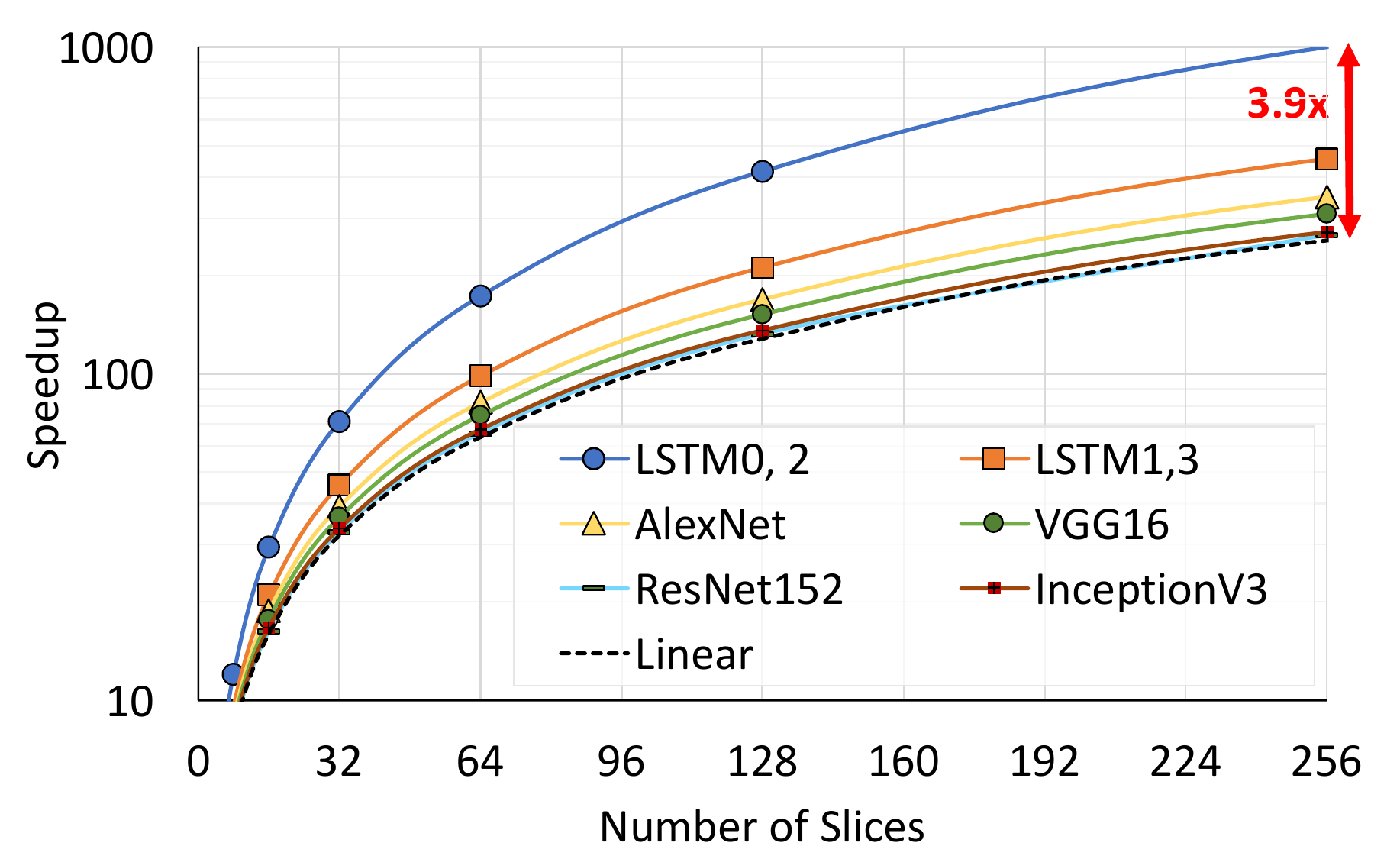}
\vspace{-10pt}
\caption{Achieving better than linear speedup when
  scaling number of slices from 2 to 256. } 
\label{scaling_all}
\vspace{-0.3in}
\end{figure}

\begin{table}[b]
\centering
\vspace{-0.3in}
\caption{Average size of weight matrices and optimal number of
  partitions that maximizes parallelism.} 
\includegraphics[width=0.7\linewidth]{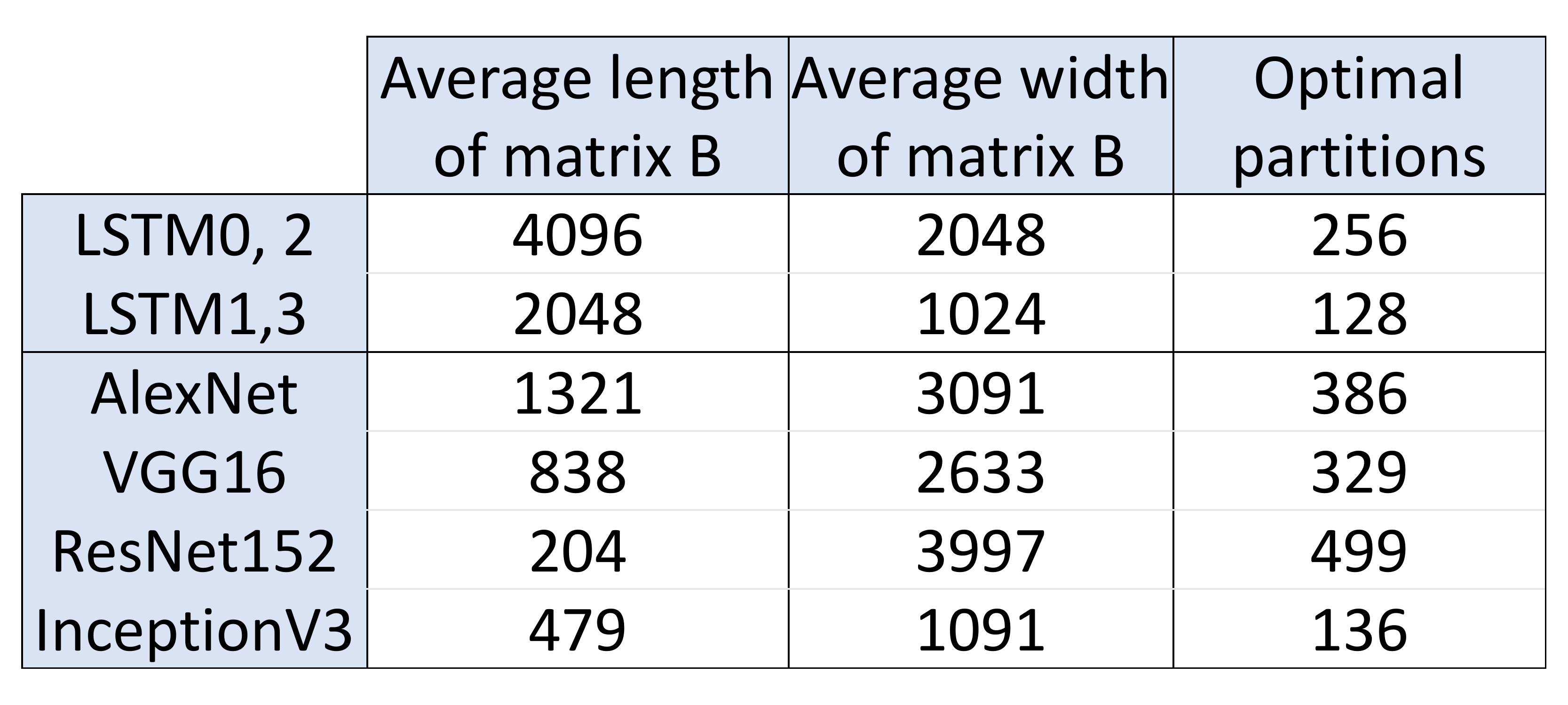} 
\label{LSTM_dimensions}
\end{table}

\subsection{Scalability}
One of our goals was to understand how we can architect NDP memory
systems where performance scales with memory size. Consequently, the
analysis in this section is mostly unique to slice-based memory
systems. There are three main outcomes from our analysis.  
First is the importance of finding the right balance between memory
bandwidth and compute bandwidth within a slice via its effect on
performance scaling.  Figure \ref{scaling} illustrates throughput for two sample
applications, one from LSTMs and one from CNNs, when the number of slices
varies from 8 to 128. For a fixed number of slices, the balanced
configuration with 2X the computation throughput/slice provides twice the
system throughput as the baseline configuration. Finding this balance
will have to come from domain analysis. The number of slices for a
specific application is addressed via memory allocation - use a subset
of slices of the complete memory system much in the same way physical
memory is allocated to processes. 

The second lesson concerns how overheads scale. As we increase the
number of slices, parallelism is exploited while overheads here have
dropped and can lead to superlinear speedup. This is primarily due
high overheads for a small number of slices and large networks as
explained below. 
While a perfect
balance at each slice (operating at the knee of the Roofline curve) is
necessary for optimum scale performance, it is not enough without
utilizing maximum parallelism across all the slices. Currently the
well-known approaches for training neural networks are data and model
parallelism\cite{dean2012large, raina2009large}, which in the best
case (i.e., when a program have enough concurrency opportunity), can
provide linear speedup\cite{TensorFlow_Bench}. On the other hand, a
system can provide better than linear speedup, if it can prevent
maintenance overhead operations, while using multiple
resources. In sum, to better scale performance, the speed up
should be gained not only from the parallelism, but also from
eliminating unnecessary operations. 

Consider when the number of
memory slices are small and we have large matrices than number of required partitions, several
parts of model and data should be processed sequentially in limited
number of slices. Handling several partitions of a layer by one slice
costs operations such as reloading weights to registers iteratively
(tasks mentioned in \ding{182} of Figure \ref{steps_micro}).  
Consider the case where the number of Memory Slices are small and we
have large matrices.  Table \ref{LSTM_dimensions} compares the weight
matrix dimensions and required partitions for utilizing all the
available fine grained parallelism of a slice-based system (when the
multiplier array width is 8). Weight matrices with a length longer
than the multiplier array length (256 in the baseline system) must be
partitioned horizontally and be loaded iteratively. In this case, when
the number of slices doubles, the latency required for those overhead
operations at each slice is reduced up to half the time, which results
up to 2.4X speedup (due cascading reductions elsewhere). Consider if
such a rate of reduction were to continue, after seven times doubling
the number of slices (e.g., from 2 to 256), the speed up will reach
$(2.4)^8/2=550$x that is 4.2 times as high as linear speedup (i.e.,
$256/2=128$). Figure \ref{scaling_all} demonstrates the speedup when
the number of slices increases from 2 to 256 (the lowest black line
shows the linear speedup trend). As the figure shows, LSTM0 and 2,
those with largest weight matrices, are better at utilizing a larger
number of slices due to larger matrix size and more parallel
operations. While generally all the eight workloads perform better on
more slices, even limited number of slices perform better than
comparable devices (see Figure \ref{GPU} and \ref{TPU}). Having to
preload the full multiplier array in a slice and 256 cycles
(and less for partial matrix partitions). While superlinear speedups
are substantively affected by this we keep in mind the usage
model. The goal is to utilize memory size to scale performance and
hence we expect larger number of slices allocated to these types of
applications when parallelism is available and envisioned memory
systems are tens to hundreds of Gbytes on a server/cloud system.

Third, slice-based memory systems benefit from the partitioning
mechanism used here. Data parallelism, in which mini-batches are split
across workers to train the same model by updating shared parameters
(the approach used in GPU clusters), has other limitations to
scalability.  
One of them is transferring large amount of data to the central server for
combining the results of SGDs.  
On completing mini-batches, error matrices are
transmitted to a central server to be combined and broadcast back to
the individual GPUs. This date grows with the size of the problem and
limits scalability. On the other hand, the approach of splitting outputs
across memory slices avoid such bottleneck by partially computing error matrices.
In addition, memory slices eliminate overheads of data parallelism, such as communication or 
storage overheads for broadcasting or duplicating the shared parameters.

\begin{figure}[t]
\centering
\includegraphics[width=1\linewidth]{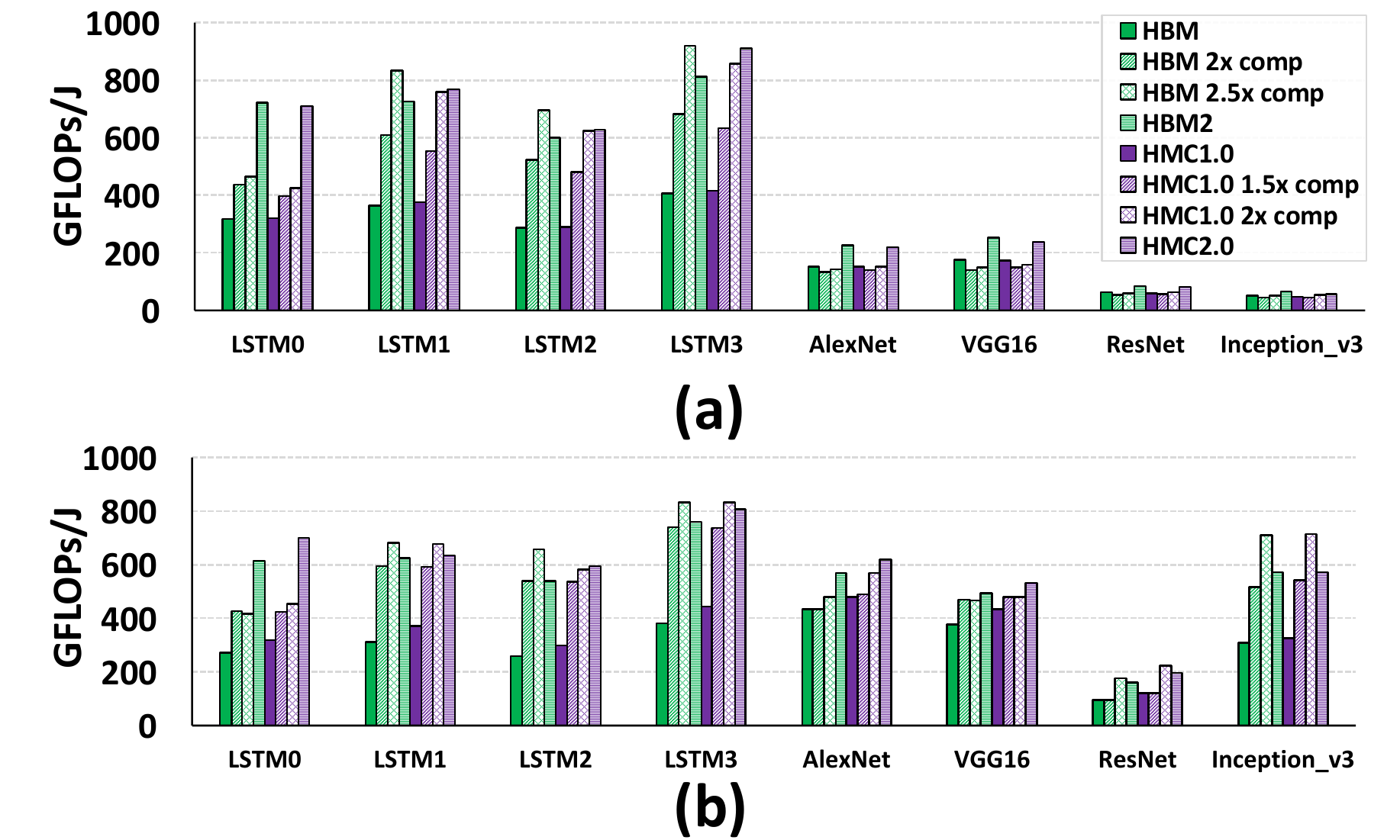}
\vspace{-20pt}
\caption{Performance efficiency of: (a) Training (b) Inference}
\label{efficiency}
\vspace{-0.2in}
\end{figure}

\subsection{Efficiency}

\begin{figure}[b]
\centering
\vspace{-0.24in}
\includegraphics[width=0.8\linewidth]{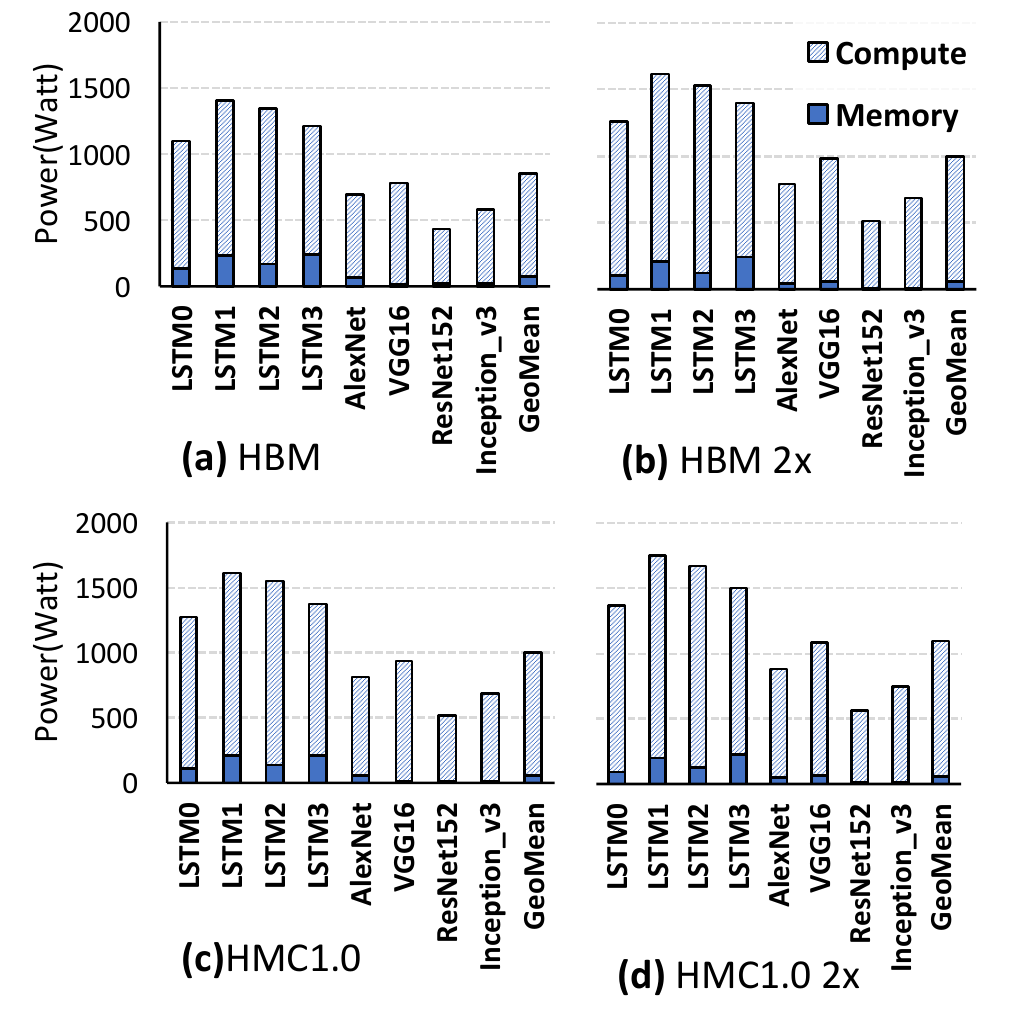}   
\vspace{-20pt}
\caption{Compute and memory power consumption of training.} 
\label{power_mem_comp}
\end{figure}

Energy consumption is an important factor in defining the cost of a
system, specifically when the power consumption budget is limited. The
number of operations that can be committed by consuming one Joule of
energy is an important metric.
Figure \ref{efficiency} compares power efficiency of training (a) and
inference (b) of workloads on various configurations. As Figure
\ref{efficiency}-a shows, those LSTMs that most utilize the
computations (i.e., LSTM1 and LSTM3), have highest performance
efficiency during training, comparing to the other LSTMs. Furthermore,
Figure \ref{power_mem_comp} illustrates a break-down of memory and
compute power consumption for training for four configurations. As can
be seen, a larger portion of power is dedicated to
computations. HBM-based configurations consumes less power because
they have a smaller number of slices to
provide the required bandwidth.  
To the best of our knowledge, the only comprehensive architectural
approach devoted to accelerating training (and for CNNs) is
ScaleDeep\cite{venkataramani2017scaledeep}, the hardware for which has
peak performance of 1.35 PFLOPs/S for half-precision(16-bit) at 1.4KW. 
A comparison between power efficiency of a slice-based memory system and
ScaleDeep shows approximately 330 GFLOPs/W for training AlexNet and
VGG16 for ScaleDeep, which is in the same range of that of the
slice-based system. 
Regarding the target of training RNNs, memory
slices can substantially improve power efficiency to 747 GFLOPs/J for LSTMs.

\section{Related Work}

\noindent 

Several studies proposed solutions for inference\cite{parashar2017scnn, gao2017tetris, kim2016neurocube} and training neural networks\cite{venkataramani2017scaledeep, azarkhish2017neurostream}.  Neurocube\cite{kim2016neurocube} that used 3D memory, TETRIS\cite{gao2017tetris} with scheduling techniques and hybrid partitioning, and SCNN\cite{parashar2017scnn} that utilizes sparsity, can be used for efficient inference of CNNs, but they do not support training dense and large-scale applications. ScaleDeep\cite{venkataramani2017scaledeep} trains CNNs by using heterogenous tile of chips for provide convolutional and fully-connected layers requirement.  
Some of the recent studies, proposed using PE arrays for neural networks. The architectures used in Eyeriss\cite{chen2017eyeriss}, DianNao\cite{chen2014dadiannao}, TETRIS\cite{gao2017tetris} devote a huge portion of each PE to SRAM memories. Therefore, they limited their PE array size to $8\times8$  up to $16\times16$.  Even Tetris, which optimizes the size of its PE array size to place it near vaults of HMC, has 512Byte register file per PE. Using a systolic array is efficient and uses only four bytes of register per compute unit (128x smaller than that for Tetris). 
TPU\cite{jouppi2017datacenter} uses a $256 \times 256$ systolic array of 8-bit integer MACs with 24MB unified buffer and 4MB of accumulator RAMs, for neural network inference. Comparing to TPU, memory slices gain higher throughput from employing systolic arrays by
 directly locating them near memory, and assigning arrays with smaller width to higher bandwidth (e.g., $8 \times 256$ to 10GB/S). In addition matched balanced to requirements of dense applications, partially adding eight-width (instead of 256) rows in parallel is more efficient.     
DianNao\cite{chen2014diannao}, DaDianNao\cite{chen2014dadiannao}, PuDianNao\cite{liu2015pudiannao}, and ShiDianNao\cite{du2015shidiannao} are  generations of neural-network accelerators, proposed for inference and training of CNNs. Their key contributions are reducing memory accesses by either utilizing NN-specific access pattern or employing eDRAM-based weight cache. Besides scalability, because of our balanced NDP design, programability, which enable training RNNs, differentiates memory slices from previous efforts.

\section{Conclusion}
Toward the growth in data-intensive applications, this paper proposed scalable and intelligent memory slices, the key microarchitectural components of which consists of a pipelined systolic-based compute logic, a programmable memory interface and an aggregation engine.  
A class of state-of-the-art compute-intensive applications including RNNs and hybrid DNNs can benefit from modularity and partitioning strategy of memory slices. 
The results of our cycle-level simulations show that memory slices provide higher throughput for inference and training of RNNs and CNNs comparing to GPUs and TPUs do. Also, memory slices exhibits a superlinear speedup when the number of slices increases.

\bibliographystyle{plain}
\bibliography{myRef}

\end{document}